\documentclass[%
 aip,
amsmath,amssymb,
reprint,%
]{revtex4-1}

\usepackage{graphicx}
\usepackage{dcolumn, multirow}
\usepackage{bm}
\usepackage[mathlines]{lineno}

\usepackage{chemformula} 
\usepackage[utf8]{inputenc}
\usepackage[T1]{fontenc}
\DeclareUnicodeCharacter{2010}{-}
\DeclareUnicodeCharacter{2009}{ }
\DeclareUnicodeCharacter{2212}{--}
\DeclareUnicodeCharacter{22C5}{.}
\DeclareUnicodeCharacter{308}{"}
\DeclareUnicodeCharacter{301}{"}

\usepackage{mathptmx}

\newcommand{\lb}{l_\mathrm{B}}
\newcommand{\kB}{k_\mathrm{B}}
\newcommand{\kt}{k_\mathrm{B}T}
\newcommand{\lin}{\mathrm{ln}}

\newcommand{\zs}{z_\mathrm{s}}
\newcommand{\Ns}{N_\mathrm{s}}

\newcommand{\cs}{c_\mathrm{s}}

\newcommand{\reff}{r_\mathrm{eff}}
\newcommand{\veff}{v_\mathrm{eff}}
\newcommand{\rd}{r_\mathrm{d}}
\newcommand{\vd}{v_\mathrm{d}}
\newcommand{\zd}{Z_\mathrm{d}}
\newcommand{\zeff}{Z_\mathrm{eff}}
\newcommand{\phieff}{\phi_\mathrm{eff}}
\newcommand{\phid}{\phi^{}_\mathrm{D}}

\newcommand{\submg}{\scalebox{.8}{$\scriptscriptstyle ++$}}
\newcommand{\subna}{\scalebox{.8}{$\scriptscriptstyle +$}}
\newcommand{\subcl}{\scalebox{.8}{$\scriptscriptstyle -$}}

\newcommand{\cbi}{c^\mathrm{m}_{i}}

\newcommand{\ctmg}{c^0_{\submg}}
\newcommand{\ctna}{c^0_{\subna}}
\newcommand{\ctcl}{c^0_{\subcl}}
\newcommand{\cti}{c^0_{i}}

\newcommand{\cmg}{c^{\mathrm{b}}_{\submg}}
\newcommand{\cna}{c^{\mathrm{b}}_{\subna}}
\newcommand{\ccl}{c^{\mathrm{b}}_{\subcl}}
\newcommand{\ci}{c^{\mathrm{b}}_{i}}

\newcommand{\ta}{\Theta_{\subna}}
\newcommand{\tb}{\Theta_{\submg}}
\newcommand{\ti}{\Theta_{i}}

\newcommand{\va}{v^0_{\subna}}
\newcommand{\vb}{v^0_{\submg}}
\newcommand{\vi}{v^0_{i}}

\newcommand{\Nb}{N_{\submg}}
\newcommand{\Na}{N_{\subna}}

\newcommand{\Ni}{N_{i}}

\newcommand{\Nbb}{N^\mathrm{b}_{\submg}}
\newcommand{\Nab}{N^\mathrm{b}_{\subna}}
\newcommand{\Nib}{N^\mathrm{b}_{i}}

\newcommand{\kaz}{\mathcal{K}_{\mathrm{int},\,{\subna}}}
\newcommand{\kbz}{\mathcal{K}_{\mathrm{int},\,{\submg}}}
\newcommand{\kcz}{\mathcal{K}_{\mathrm{int},\,{\subcl}}}
\newcommand{\kiz}{\mathcal{K}_{\mathrm{int},\,{i}}}
\newcommand{\kiel}{\mathcal{K}_{\mathrm{el},\,{i}}}

\newcommand{\ki}{\mathcal{K}_{i}}

\newcommand{\dg}[1]{\Delta \mu_{\mathrm{int},\,{#1}}}
\newcommand{\dgb}{\Delta \mu_{\mathrm{int},\,{\submg}}}
\newcommand{\dga}{\Delta \mu_{\mathrm{int},\,{\subna}}}

\newcommand{\dgi}{\Delta \mu_{\mathrm{int},\,i}}
\newcommand{\dgiel}{\Delta \mu_{\mathrm{el},\,i}}
\newcommand{\dgicr}{\Delta \mu_{\mathrm{tr},\,i}}
\newcommand{\dgimx}{\Delta \mu_{\mathrm{mix},\,i}}

\newcommand{\dgel}{\Delta \mu_\mathrm{el}}
\newcommand{\dgcr}{\Delta \mu_\mathrm{tr}}
\newcommand{\dgmx}{\Delta \mu_\mathrm{mix}}

\newcommand{\dGi}{\Delta \mu_{\mathrm{bind},\,i}}

\newcommand{\fel}{\mathcal{F}_\mathrm{el}}
\newcommand{\fmx}{\mathcal{F}_\mathrm{mix}}
\newcommand{\fid}{\mathcal{F}_\mathrm{tr}}
\newcommand{\fin}{\mathcal{F}_\mathrm{int}}
\newcommand{\ft}{\mathcal{F}_\mathrm{tot}}

\newcommand{\dd}{\mathop{}\!\mathrm{d}}

\newcolumntype{P}[1]{>{\centering\arraybackslash}p{#1}}
\newcolumntype{L}{>{$}l<{$}} 
\usepackage{dcolumn} 

\begin{document}

\preprint{AIP/123-QED}

\title{Competitive sorption of mono- versus divalent ions by highly charged globular macromolecules}

\author{Rohit Nikam}
\affiliation{Research Group for Simulations of Energy Materials, Helmholtz-Zentrum Berlin f{\"u}r Materialien und Energie, Hahn-Meitner-Platz 1, D-14109 Berlin, Germany}
\affiliation{Institut f{\"u}r Physik, Humboldt-Universit{\"a}t zu Berlin, Newtonstr.~15, D-12489 Berlin, Germany}
\author{Xiao Xu}
\affiliation{School of Chemical Engineering, Nanjing University of Science and Technology, 200 Xiao Ling Wei, Nanjing 210094, P. R. China}
\author{Matej Kandu\v{c}}
\affiliation{Department of Theoretical Physics, Jo\v{z}ef Stefan Institute, Jamova 39, SI-1000 Ljubljana, Slovenia}
\author{Joachim Dzubiella}
\affiliation{Research Group for Simulations of Energy Materials, Helmholtz-Zentrum Berlin f{\"u}r Materialien und Energie, Hahn-Meitner-Platz 1, D-14109 Berlin, Germany}
\affiliation{Applied Theoretical Physics -- Computational Physics, Physikalisches Institut, Albert-Ludwigs-Universit{\"a}t Freiburg, Hermann-Herder-Str.~3, D-79104 Freiburg, Germany}
\email{joachim.dzubiella@physik.uni-freiburg.de}

\begin{abstract} 
When a highly charged globular macromolecule, such as a dendritic polyelectrolyte or charged nanogel, is immersed into a physiological electrolyte solution, monovalent and divalent counterions from the solution bind to the macromolecule in a certain ratio and thereby almost completely electroneutralize it. For charged macromolecules in biological media, the number ratio of bound mono- versus divalent ions is decisive for the desired function. A theoretical prediction of such a sorption ratio is challenging because of the competition of electrostatic (valency), ion-specific, and binding saturation effects. Here, we devise and discuss a few approximate models to predict such an equilibrium sorption ratio by extending and combining established electrostatic binding theories such as Donnan, Langmuir, Manning as well as Poisson--Boltzmann approaches, to systematically study the competitive uptake of mono- and divalent counterions by the macromolecule. We compare and fit our models to coarse-grained (implicit-solvent) computer simulation data of the globular polyelectrolyte dendritic polyglycerol sulfate (dPGS) in salt solutions of mixed valencies. The dPGS has high potential to serve in macromolecular carrier applications in biological systems and at the same time constitutes a good model system for a highly charged macromolecule. We finally use the simulation-informed models to extrapolate and predict electrostatic features such as the effective charge as a function of the divalent ion concentration for a wide range of dPGS generations (sizes). 
\end{abstract}

\maketitle

\section{\label{intro} Introduction} 
Polyelectrolytes in polar solvents such as water are important and ubiquitous in biological as well as in synthetic matter.~\cite{Muthukumar2017,Katchalsky1964,Rubinstein2012,Boroudjerdi2005,Forster1995,Dobrynin2005,Liu2003}
In these systems, electrostatic interactions, regulated by free ions and water, play a dominant role in shaping the structural and electrostatic characteristics of the polyelectrolyte, and the subsequent function of the system.~\cite{Muthukumar2017,Rubinstein2012,Boroudjerdi2005}
The electrostatic attraction between the isolated polyelectrolyte molecule and the oppositely charged counterions in the solution leads to strong counterion condensation on the molecule. 
This significantly modifies its interaction with other charged molecules (\textit{e.g.}, proteins, DNA, etc.) and its electric properties such as the electrophoretic mobility in an external electric field.~\cite{Boroudjerdi2005,Forster1995,Liu2003}
Therefore, understanding counterion condensation is of utmost importance in order to understand the properties of polyelectrolytes and their implications in the biological and synthetic environments.~\cite{Chremos2016,Boroudjerdi2005} Condensation effectively leads to neutralizing an equivalent amount of the structural charge $\zd$ of the macromolecule.~\cite{alexander1984charge, belloni1998ionic} Hence, the charged substrate plus its confined counterions may be considered as a single entity with an effective (or renormalized) charge $\zeff$, which is significantly lower than the bare structural charge $\zd$. One can then identify the difference $\zd-\zeff$ as the amount of counterions condensed in the surface region.~\cite{Bocquet2002}

The phenomenon of counterion condensation and the effect of ionic strength on the configurational properties of different types of polyelectrolyte molecules such as chains,~\cite{Forster1992,dobrynin1995,Wenner2002,Raspaud1998,Dobrynin2005,Liu2002,Liu2003,Muthukumar2004,Chremos2016} brushes,~\cite{ruhe2004polyelectrolyte,pincus1991colloid,borisov1991collapse,Zhulina1995,Zhulina2000} or polyelectrolyte nanogels~\cite{nanogel1,nanogel2,nanogel3,arturo2017,arturo2018} have been studied extensively in the past. Through the knowledge of the distribution of the salt ions around the polyelectrolyte, \textit{e.g.}, measured in terms of the radial distribution function in simulations and experiments, it is possible to derive important properties such as charge--charge correlation, osmotic compressibility and shear viscosity of the system.~\cite{forster1995polyelectrolytes}
Muthukumar, in his extensive and comprehensive review of the experimental, theoretical and simulation based research done on polyelectrolyte chains, described the effect of salt concentration, valency of counterions, chain length and polyelectrolyte concentration on counterion condensation.~\cite{Muthukumar2017,manning2012poisson}
Besides the properties of a single isolated polyelectrolyte molecule, the ionic strength of the solution also influences the interaction of polyelectrolytes with other entities, such as adsorption on substrates,~\cite{VandeSteeg1992,Dahlgren1993,Netz1999,Hariharan1998,Gittins2001,caruso2000hollow} formation of ultra-thin polyelectrolyte multilayer membranes,~\cite{Decher1992,Decher1997,Ladam2000,McAloney2001,Dubas2001} the structure and solubility of polyelectrolyte complexes~\cite{hugerth1997effect,rusu2003formation,winkler2002complex,Kudlay2004a,Mende2002} or coacervates.~\cite{spruijt2010binodal,Gucht2011,biesheuvel2004electrostatic,Perry2014} 

As an emerging class of functional polyelectrolytes, polyelectrolyte nanogels~\cite{nanogel1,nanogel2,nanogel3,arturo2017,arturo2018} and dendritic or hyperbranched polyelectrolytes~\cite{JensDernedde2010,Khandare2012, Groeger2013, Maysinger2015,Reimann2015} have attracted considerable interest in the scientific community in the last years due to their multifaceted bioapplications, such as biological imaging, drug delivery and tissue engineering.~\cite{Leereview, Ballauff2004,Tian2013}  In particular, the hyperbranched or dendritic polyglycerol sulfate molecules (hPGS or dPGS, respectively) are found to possess strong anti-inflammatory properties,\cite{Maysinger2015, Reimann2015} act as a transport vehicle for drugs towards tumor cells,\cite{Sousa-Herves2015, Groeger2013, Vonnemann2014} and can be used as imaging agents for the diagnosis of rheumatoid arthritis.~\cite{Vonnemann2014} This wide variety of applications, thus, have proven them to be high potential candidates for the use in medical treatments.~\cite{Khandare2012} Hence, the understanding of dPGS interaction with the \textit{in vivo} environment becomes important. The highly symmetric dendritic topology, terminated with monovalent negatively charged sulfate groups, makes dPGS also an excellent representative model in the class of highly charged globular polyelectrolytes.~\cite{xu2017charged, nikam2018charge} Because of the charged terminal groups, dPGS mainly interacts through electrostatics, rendering counterion condensation and subsequent charge renormalization effects to become substantial for function.  

There have been past efforts to investigate the counterion condensation and to define the effective charge as a result of the charge renormalization on charged hard-sphere colloids.~\cite{Ohshima1982, Zimm1983, alexander1984charge, Belloni1984, belloni1998ionic, Ramanath1988, Manning2007, Bocquet2002, Gillespie2014} 
However, the characterization of open-structure nanogel particles or dendrites such dPGS, which in part are penetrable to ions and a surface is not well defined, remains challenging.~\cite{Ohshima2008} 
Recently, Xu~\textit{et al.} implemented a simple but accurate scheme to define and determine the effective surface potential and its location for dPGS, by mapping potentials obtained from simulations to the Debye--H\"uckel potential in the far-field regime.~\cite{xu2017charged} 
This scheme is widely known as the Alexander prescription.~\cite{alexander1984charge,Trizac2002,bocquet2002effective,Levin2004}  Based on this criterion, a systematic electrostatic characterization of dPGS has been performed via coarse-grained~\cite{xu2017charged} and all-atom~\cite{nikam2018charge} simulations by defining the number of condensed (bound) ions.  It was then established that the strong binding of dPGS to lysozyme -- an abundant protein in the human biological environment -- and a sequential formation of a protein corona around dPGS in the presence of NaCl salt solution, is dominantly governed by the entropic gain due to the release of a few Na$^+$ counterions during binding.~\cite{xu:biomacro} 
Proteins typically bind strongly to the macromolecular surface, thereby forming a protein `corona', a dense shell of proteins that can entirely coat the macromolecule.~\cite{Owens2006, Cedervall2007, Lindman2007, Monopoli2012, wang2013biomolecular, LoGiudice2016, Boselli2017}

Considering the medicinal applications of dPGS, it is important to study its interactions with  {\it divalent metal cations}, \textit{viz.} magnesium(II) and calcium(II) ions,  which are key constituents of the human blood serum. 
Mg$^{2+}$ is essential for the stabilization of proteins, polysaccharides, lipids and DNA/RNA molecules, while Ca$^{2+}$ is critical for bone formation and plays a key role in signal transduction.~\cite{Friesen2019a, da2001biological} 
Human serum blood contains approximately $0.75-0.95$\,mM Mg$^{2+}$ ions, $1-4$\,mM Ca$^{2+}$ ions and around $150$\,mM NaCl salt in a dissociated form.~\cite{meyers2004encyclopedia, kretsinger2013encyclopedia} 
Thus, upon the administration of dPGS into the human biological environment, it is imperative for the competitive adsorption between the divalent (Mg$^{2+}$/Ca$^{2+}$) and monovalent (Na$^+$) ions to establish on the dPGS molecule, which can change the effective charge, and subsequently the interaction properties of dPGS with other charged entities such as proteins. 
This microscopic mechanism has a potential to significantly alter the attributes of protein corona around dPGS,~\cite{XiaoCPS} thus, the biological immune response to the dPGS--protein corona complex, its metabolic fate, and the function of such a complex in biomedical and biotechnological applications.
The competitive ion binding can be observed also in a wide variety of the biological and industrial ion-exchange processes such as the alkaline-earth/alkali-metal ion-exchange onto polyelectrolytes,~\cite{Pochard1999} desalination of saline water to produce potable water,~\cite{Birnhack2019} demineralization of whey, acid and alkali recovery from waste acid~\cite{kobuchi1986application} and alkali solutions~\cite{sata1993new} by diffusion dialysis,~\cite{Sata2002} etc.

Interactions of multivalent ions with polyelectrolyte solutions have been theoretically studied in the past, in terms of their thermodynamic properties,~\cite{Kuhn1999} ionic and potential distributions,~\cite{Gavryushov1997} accurate calculation of the effective charge,~\cite{DosSantos2010} and the effect on the interaction between polyelectrolyte macromolecules.~\cite{Arenzon1999,Naji2004,Kanduc2010,Rudi2016}
In this paper, the focus is to theoretically analyze the competitive sorption of {\it mono- versus divalent counterions} by highly charged spherical dPGS-like polyelectrolytes with the help of mean-field continuum and discrete binding site models, informed by coarse-grained computer simulations of dPGS of various generations.  The theoretical models are generally formulated for globular charged macromolecules and include ion-specific effects in a parametric way and can thus be straightforwardly modified or adapted to other charged globules, where mono-/divalent ion-exchange plays a role.  In particular, we begin with the simple Donnan model, modified for ion-specific uptake, assuming that the electrostatic potential and the ionic concentrations are constant within the macromolecule phase and the bulk phase.~\cite{Basser1993, arturo2016, Ahualli2014} Despite being simple, still, for the mixed case of monovalent and divalent ions the resultant composition is a non-trivial outcome.  We continue with the mean-field Poisson--Boltzmann (PB) model, widely used in colloidal science and electrochemistry,~\cite{Rubinstein2012,israelachvili2011intermolecular, adamson1967physical, verwey1947theory, Borukhov2000} and with the limitations well known and discussed, in particular the neglect of electrostatic and steric correlations,~\cite{eigen1954, kralj1996, Cuvillier1997, Borukhov1997,DosSantos2010} or ion-specific sorption effects.~\cite{Kalcher2010,Kalcher2010a,Chudoba2018,arturo2014,Yaakov2009a,Lima2008,Koelsch2007,okur2017beyond,LoNostro2012,Schwierz2013}  
The PB model has also been implemented to address the problem of competitive counterion binding in a mixed salt for the cases of linear polyelectrolytes such as DNA~\cite{Burak2004,Chen2002,Rouzina1997,Misra1994,Paulsen1987} and planar geometries.~\cite{Rouzina1994}
We also devise a two-state approximation model for an ion condensation around a charged globule. 
\begin{figure*}[ht!]
\centering
\includegraphics[width=14cm, height=14cm, keepaspectratio]{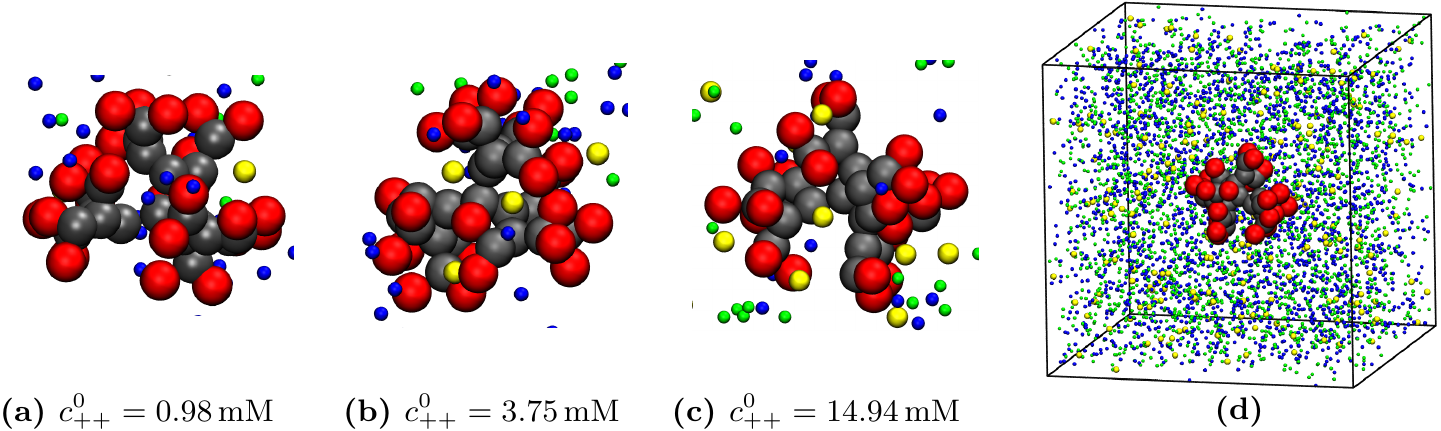}
\caption{\label{snap} Panels (a), (b) and (c) show coarse-grained (CG) simulation snapshots of G$_2$-dPGS in a mixture of ions at the divalent cation (DC) concentrations of $0.98$\,mM, $3.75$\,mM and $14.94$\,mM, respectively, while the monovalent salt concentration $c^0_{+}$ is at $150.37$\,mM.  The red beads depict the charged terminal sulfate groups (--\ch{OSO3}), which represent the binding sites of dPGS, the gray beads depict the neutral glycerol (\ch{C3H5O}--) branching units, and yellow, blue and green spheres refer to DCs, monovalent cations (MCs) and monovalent anions, respectively. 
(d) Snapshot of the whole simulation box containing the CG model of G$_2$-dPGS and a mixture of salts of MCs and DCs. The box is cubic with a side length $L=30$~nm.
The sizes of spheres/beads in all panels are not to scale. }
\end{figure*}
The two-state approach was firstly used in the Oosawa--Manning model~\cite{Manning1969, Oosawa} for the counterion condensation around polyelectrolyte chains, according to which, counterions in a solution can be classified into two categories: `free' counterions, which are able to explore the whole solution volume $V$ and the `condensed' (or `bound') counterions, which are localized within a small volume surrounding the polyelectrolyte macromolecule. An equivalent model for an impenetrable sphere with a surface charge was developed by Manning, where the number of condensed counterions on the macromolecule per bare unit surface charge is obtained by a free energy minimization, pointing to the competition between the electrostatic binding of counterions to the macromolecule and their dissociation entropy.~\cite{Manning2007}  We extend this model by introducing a discrete binding site model by considering the finite configurational volume of the ion in the condensed state and that the macromolecule has a finite number of charged binding sites by adopting the mixing entropy from the works of McGhee and von Hippel.~\cite{Mcghee1974} 
Ion-binding models in the same spirit have been developed in the past to describe the ionization equilibrium of {\it linear} polyelectrolytes in monovalent salt~\cite{flory1953molecular,Raphael1990,Muthukumar2004}, multivalent salt,~\cite{Friedman1984} and in mixtures of mono- and divalent salts.~\cite{Kundagrami2008} All our models are compared to molecular simulations and used to study systematically the key electrostatic features of a highly charged globule, such as the effect of competitive adsorption on the variation of the number of condensed monovalent and divalent counterions, effective charge, and its variation with divalent ion concentration.

\section{Coarse-grained computer simulations}\label{cg_sim}
\subsection{Simulation methods, force fields, and systems}\label{sec:cg_ff}
The coarse-grained (CG) monomer-resolved models of the dPGS macromolecule have been developed previously~\cite{xu2017charged} and maintain the essential dPGS structural and electrostatic features with affordable computing expense. 
In brief, the dPGS branching units (\ch{C3H5O}--) and inner core (\ch{C3H5}--) (both of which are a part of the glycerol chemical group, respectively), and the terminal sulfate groups (--\ch{OSO3}) are individually represented by the CG segments of specific type. 
The gross number of the CG segments is equal to the dendrimer polymerization $N_g =  3 \times 2^{n+1} - 2$ of generation index $n$. 
Only the terminal segments are charged with $-1e$ (where $e$ is the elementary charge), leading to the dPGS bare valency $|Z_n| = 3 \times 2^{n+1}$.
The CG segments are connected by bonded and angular potentials both in harmonic form. In the previous work \cite{xu2017charged} we only studied monovalent ions. Here we extend it to study the competitive uptake of mono- and divalent ions for generations 2 and 4. The bare charge valencies of the G$_2$-dPGS are thus $\zd = Z_{n=2} = -24$ and $\zd = Z_{n=4} = -96$. Snapshots are shown in Fig.~\ref{snap}. 
 
The non-bonded interactions between CG beads are described by the Lennard-Jones (LJ) potential together with the Lorentz--Berthelot mixing rules. In particular, the energy parameter $\epsilon_{\rm LJ} = 0.1\,\kt$ and the diameter $\sigma_{\rm LJ} = 0.4$~nm are set identical 
for all ions (mono- and divalent) and thus any ion-specific effects are not explicitly included. 
In our simulations we place the dPGS in the center of a periodically repeated cubic box with a volume of $V$ (side-length of $L = 30$~nm). 
The solvent is implicitly assumed as a dielectric continuum with a dielectric constant $\epsilon_{\rm w} = 78$.  The CG simulations employ the stochastic dynamics (SD) integrator in Gromacs $4.5.5$ as in our previous work.~\cite{xu2017charged}

\begin{figure*}[htbp!]
\centering
\includegraphics[width=12.5cm, height=12.5cm, keepaspectratio]{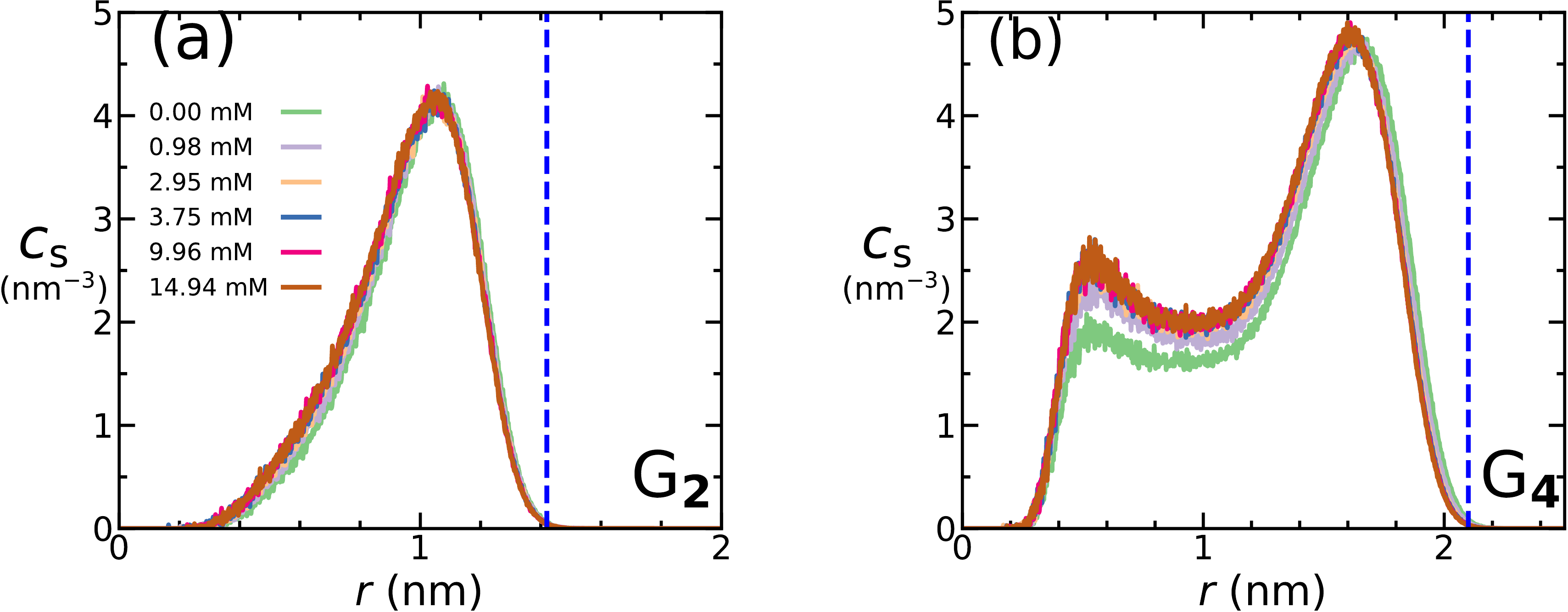}
\caption{\label{rho_sulfate} Radial density distributions $\cs(r)$ of the terminal sulfate groups of dPGS as a function of the distance $r$ from the COM of (a) G$_2$-dPGS and (b) G$_4$-dPGS, obtained from the coarse-grained computer simulations. 
The curves are plotted for different DC concentrations $c^0_{++}$ (see legend).
The blue vertical dashed lines denote the dPGS bare radius $\rd$ ($1.4$\,nm for G$_2$-dPGS and $2.1$\,nm for G$_4$-dPGS) defined as the location where $\cs(r)$ falls to the physiological threshold of $150$\,mM.}
\end{figure*}

\begin{figure*}[ht!]
\centering
\includegraphics[width=13cm, height=13cm, keepaspectratio]{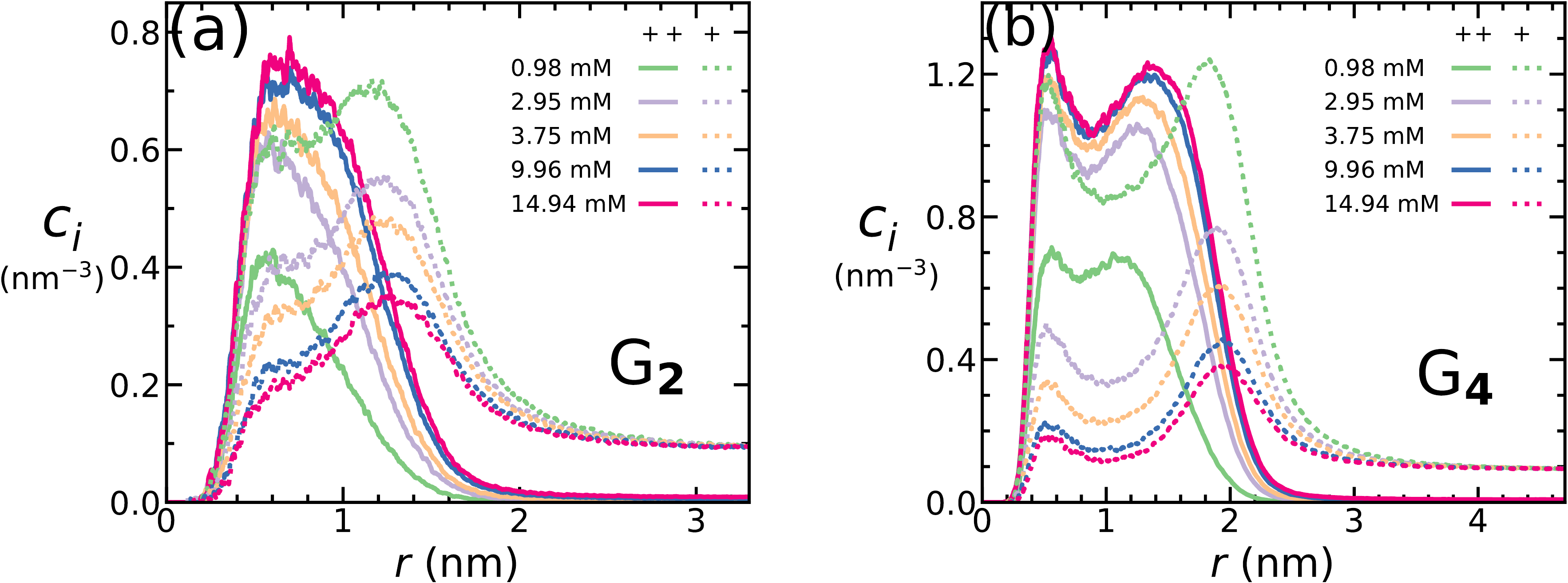}
\caption{\label{density_ions} Radial density distributions $c_i(r)$ ($i=+,+\!+$) of counterion species $i$ as a function of the distance $r$ from the COM of (a) G$_2$-dPGS and (b) G$_4$-dPGS. 
The curves are plotted for different DC concentrations $c^0_{++}$ (see legend).
The solid and dotted lines depict the density distributions of DCs and MCs, respectively.
}
\end{figure*} 

All simulations are performed in the canonical ensemble. The divalent cations (DCs), monovalent cations (MCs) and monovalent anions in the system are referred to with subscripts $++,\; +$ and $-$, respectively.  The dPGS is accompanied by the corresponding number of monovalent counterions $\Ns$ ($24$ for G$_2$-dPGS and $96$ for G$_4$-dPGS) electrically neutralizing the macromolecule and having the same chemical identity as the MCs of the salt. The number of salt ions $i$ ($i = +\!+,+,-$) is denoted as $n_i$, while the corresponding total salt concentrations are denoted as $\cti=n_i/V$. Bulk concentrations are defined as $\ci=(n_i - \Nib)/(V - \veff)$ (for $i=+\!+,-$) and $\cna=(n_{\subna} +\Ns - \Nab)/(V - \veff)$, where $\veff= 4\pi \reff^3/3$ is the volume enclosed by the effective radius $\reff$ of dPGS and $\Nib$ is the number of ions $i$ condensed (bound) on the dPGS. The definitions of both $\reff$ and $\Nib$ are discussed in Section~\ref{sim_analysis}. 

The simulations are performed at the total DC concentrations $\ctmg$ of 0.98, 2.95, 3.75, 9.96 and 14.94\,mM. G$_2$-dPGS simulation snapshots for different $\ctmg$ values are shown in Fig.~\ref{snap}(a)-(c), while the whole simulation box is displayed in Fig.~\ref{snap}(d). 
The MC concentration $\ctna$ is fixed to $150.37$\,mM and the monovalent anion concentration is adjusted in a way to ensure electroneutrality in the simulation box. 
The bulk ionic strength $I = \frac{1}{2}\sum_i z^2_i \ci$ ($i = +,+\!+,-$ with the charge valency $z_i$) ranges from $150.5$\,mM to $195$\,mM. 
The corresponding Debye screening length $\kappa^{-1} = \left(8\pi \lb I \right)^{-1/2}$ (where $\lb$ is the Bjerrum length)
ranges from $0.8$\,nm  ($\cmg=0$ and $\cna=150.5$\,mM) to $0.7$\,nm ($\cmg=14.94$ and $\cna=150.5$\,mM). 
As a reference, we also perform CG simulations in the limit of only monovalent salt, with total concentrations $\ctna$ of 10.02, 25.06 and 150.37\,mM. 

\subsection{Simulation results: radial density distributions}\label{sim_results}

The dPGS structure and its response to the addition of the DCs, is examined by the density distribution of the terminal sulfate beads $\cs(r)$ as a function of the distance $r$ from the center-of-mass (COM) of the dPGS, for different DC concentrations $\ctmg$, as shown in Fig.~\ref{rho_sulfate}. Interestingly, the presence of DCs does not lead to a notable change in the dPGS structure. 
Instead, the $\cs(r)$ profiles in the operated range of $\ctmg$ and for both G$_2$-dPGS and G$_4$-dPGS are reasonably coincident. 
Fig.~\ref{rho_sulfate}(a) shows that for G$_2$-dPGS, a single-peak distribution is found, indicating that most of the sulfate beads reside on the molecular surface.
However, in Fig.~\ref{rho_sulfate}(b), a bimodal distribution is seen for G$_4$-dPGS with a small peak at $r \simeq 0.6$\,nm. 
This backfolding phenomenon, contributing to a dense-core arrangement due to the dense macromolecular shell,~\cite{Ballauff2004} is also found in our previous works~\cite{xu2017charged,nikam2018charge} and has been detected for other terminally charged CG dendrimer models.~\cite{Huismann2010,Huismann2010B,Klos2010,Klos2011}
After the major peak, $\cs(r)$ gradually subsides to zero. 
The location where the charge density $\cs(r)$ falls to $150$\,mM, which we set as the physiological NaCl concentration, is defined as the bare (intrinsic) radius of dPGS $\rd$,~\footnote{$\rd$ in our previous works is defined as the location of the major peak of the sulfate density distribution.~\cite{xu2017charged,nikam2018charge}} shown as vertical dashed blue lines in Fig.~\ref{rho_sulfate}. 
The $\rd$ values for G$_2$-dPGS and G$_4$-dPGS are obtained as $1.40$\,nm and $2.11$\,nm, respectively.
Fig.~\ref{rho_sulfate}(b) also shows that a slight shift in the location of the major peak and an enrichment of the lower peak appears as $\ctmg$ increases, indicating a slow shrinking of the dPGS molecule due to the condensation of DCs (see Fig.~\ref{density_ions}). 

Figs.~\ref{density_ions}(a) and (b) show the cation density distributions $c_i(r)$ ($i =+,+\!+$) for  G$_2$-dPGS and G$_4$-dPGS, respectively. 
Let us focus first on G$_2$ in Fig.~\ref{density_ions}(a). 
The MC distribution $c_{\subna}(r)$ shows a high accumulation of counterions close to the sulfate groups, with a global maximum at distances $r \sim 1.2$\,nm, slightly larger than the sulfate peak (peaking roughly at $\sim 1$\,nm). 
This means that the most strongly bound `condensed' MCs are thus distributed more on the surface layers of the dPGS. 
At larger distances, $r \sim 2$\,nm, a Debye--H\"uckel like decay is observed. 
Adding more DCs, the MC distribution gradually diminishes, as expected from the exchange of MCs with DCs within the dPGS. 
However, interestingly, the DC distribution peaks at distances distinctively smaller than the location of the sulfate peak, roughly $0.5-0.6$\,nm shifted towards the dPGS center away from the peak of the MC distribution. 
This more interior binding might be attributed to different binding mechanisms between DCs and sulfate, \textit{e.g.}, bridging of two sulfate groups by one DC, which might be sterically favored closer to the dPGS core. 
These subtle structural effects may have important consequences in the context of the counterion-release mechanism driving the dPGS--protein binding,~\cite{xu:biomacro} which should be interesting for future studies. 
The ion profiles for G$_4$ shown in Fig.~\ref{density_ions}(b) show qualitatively the same behavior but are broader and double-peaked because of the significant sulfate backfolding as previously presented in Fig.~\ref{rho_sulfate}(b).

It is worth noting that simulations of DCs in general are more challenging than for MCs only. DC are more heavily hydrated than MCs (\textit{e.g.}, Mg$^{2+}$ and Na$^+$ ions),~\cite{Stokes1948,Marcus2006} therefore future studies should scrutinize the ionic size used in the implicit solvent.  
Furthermore, quantum mechanical charge transfer effects as a result of the ion-induced powerful electronic polarization of the surrounding media,~\cite{Yao2015} which are much more prevalent in the case of DCs~\cite{Pavlov1998,Kohagen2014} than MCs, may also be subsumed in ionic sizes  in the implicit water.  These model details may subtly change the density profiles shown in Figs.~\ref{density_ions}(a) and (b). However, the effects on total competitive uptake should be relatively minor as they are dominantly driven by valency and electrostatic correlations, and size effects are typically of second order importance.

Using the density distributions of the charged entities shown above, the electrostatic properties of dPGS can be studied in the presence of the mixture of DCs and MCs. The analysis methods described in Section~\ref{sim_analysis} are used to define the effective radius $\reff$, charge valency $\zeff$ and potential $\phieff$ of dPGS.  

\subsection{Electrostatic properties of dPGS}\label{sim_analysis}

\textbf{dPGS effective radius} The first step to study the ion condensation behavior is to adopt a characteristic distance $\reff$ to distinguish a bound ion from an unbound one. A practical method in that respect has been summarized in our previous work.~\cite{xu2017charged}
In short, we first consider the dPGS radial electrostatic potential profile $\phi$ (scaled by $\kt/e$), through the framework of the Poisson's equation
\begin{equation}
\nabla^2 \phi = -4\pi \lb \sum_{i} z_i c_i(r) \qquad  i= s, ++, +, -
\label{possion}
\end{equation}
Here, $c_i(r)$ refers to the radial number density profiles with respect to the distance to the dPGS-COM $r$ for all charged species in the CG simulation, 
namely, sulfates ($s$), DCs, MCs, and monovalent anions. For all ionic species, $c_i(r)$ reaches the bulk number density $\ci$ in the far-field. The simulation results for the profiles are shown in Figs.~\ref{rho_sulfate} and \ref{density_ions}.
The Poisson's equation is numerically integrated twice to obtain $\phi(r)$, which is then compared with the dimensionless Debye--H\"uckel potential $\phi^{}_\mathrm{DH}$, given by~\cite{alexander1984charge,xu2017charged,nikam2018charge}
\begin{equation}
\phi^{}_\mathrm{DH}(r) = \zeff \lb \frac{\mathrm{e}^{\kappa \reff}}{1+\kappa \reff} \frac{\mathrm{e}^{-\kappa r}}{r}.
\label{DH_eq}
\end{equation}
$\phi^{}_{\rm DH}$ is applicable to a charged sphere with radius $\reff$ and valency $\zeff$.
It approaches to $\phi$ only after the distance $r^*$ where non-linear effects, including the correlation and condensation of ions, subside. Thus, $r^* = \reff$ is eligible to serve as the dPGS effective radius to define the bound ions. 
The effective surface potential of dPGS obtained from simulations is then defined as $\phieff = \phi(\reff)$, which is shown in Table~\ref{cg_table}.
Comparing Eq.~\eqref{DH_eq} to the radial electrostatic potentials from the simulations,~\cite{xu2017charged} the value of $\reff$ for dPGS in the simulations for G$_2$ and G$_4$ was found to be $1.65$\,nm and $2.40$\,nm, respectively, under the operated concentration range in the mixture of DCs with MCs as well as in the monovalent limit, as shown in Table~\ref{cg_table}. 
These values are different than the ones obtained in our previous work,~\cite{xu2017charged} which operates at $\ctna = 10$\,mM, unlike the current work where $\ctna = 150.37$\,mM.
The newly obtained $\reff$ values in this work are then used as an input for the MMvH model, as discussed in Section~\ref{mvh}, to describe the competitive sorption. 
It is thus implicitly assumed that $\reff$ does not depend on the sorption of DCs, within the operated range of DC concentrations. 
The same prescription will be used to define $\reff$ (denoted as $\reff^\mathrm{PB}$) from the solutions of the PPB model, as discussed in Section~\ref{sec:ppbl}. 
The results for $\reff^\mathrm{PB}$ are also shown in Table~\ref{cg_table}.

\begin{table*}[ht!]
\caption{ \label{cg_table} Structural and electrostatic parameters of G$_2$-dPGS (having the bare charge valency $\zd = -24$) and G$_4$-dPGS ($\zd = -96$) measured from the coarse-grained (CG) simulations and according to ion-specific penetrable Poisson--Boltzmann (PPB) model (described in Section~\ref{sec:ppbl}). $r_{\rm eff}$, $Z_{\rm eff}$, $\phi_{\rm eff}$ are the effective radii, charge valency and potential of the dPGS, respectively, as a function of the DC concentration $c^0_{++}$, evaluated via simulations. 
The simulation box is cubic with a side length of 30~nm.
$r^\mathrm{PB}_{\rm eff}$, $Z^\mathrm{PB}_{\rm eff}$, $\phi^\mathrm{PB}_{\rm eff}$ are the effective radii, charge valency and potential of the dPGS, respectively, calculated via the PPB model. The MC concentration $c^0_{+}$ is set to $150.37$~mM. }
\setlength{\tabcolsep}{4pt} 
\renewcommand{\arraystretch}{1.3}
\begin{tabular}{c|c c c c | c c c c|| c c c | c c c} \hline \hline
  &  \multicolumn{8}{c||}{CG Simulation} & \multicolumn{6}{c}{PPB}  \\ \cline{2-15}
  &  \multicolumn{4}{c|}{G$_2$}  & \multicolumn{4}{c||}{G$_4$} &  \multicolumn{3}{c|}{G$_2$}  & \multicolumn{3}{c}{G$_4$}  \\ \cline{2-15}
$\ctmg$ & $\rd$ & $\reff$  & $\zeff$ & $\phieff$ & $\rd$ & $\reff$ & $\zeff$ & $\phieff$  & $\reff^\mathrm{PB}$  & $\zeff^\mathrm{PB}$ & $\phieff^\mathrm{PB}$ & $\reff^\mathrm{PB}$  & $\zeff^\mathrm{PB}$ & $\phieff^\mathrm{PB}$ \\
({\rm mM})  & ({\rm nm}) & ({\rm nm}) &  &  & ({\rm nm})  & ({\rm nm}) &   &  & ({\rm nm}) &  &  & ({\rm nm}) & &  \\ 
$0.00$  & \multirow{6}{*}{$1.41$} & \multirow{6}{*}{$1.65$} & $-10.09$  & $-1.26$ & \multirow{6}{*}{$2.11$} & \multirow{6}{*}{$2.40$} & $-20.04$  & $-1.27$  & \multirow{6}{*}{$1.42$} & $-11.72$  & $ -1.32$ & \multirow{6}{*}{$2.36$} & $-23.60$  & $-1.56$ \\
$0.98$  &  & & $-8.85$ & $-1.15$ &  & & $-17.75$ & $-1.14$ & & $-9.79$ & $-1.12$ & & $-20.03$ & $-1.38$ \\ 
$2.95$  &  & & $-7.40$ & $-0.98$ &  & & $-14.21$ & $-0.93$ & & $-8.89$ & $-0.88$ & & $-15.54$ & $-1.05$ \\
$3.75$  &  & & $-6.84$ & $-0.85$ &  & & $-12.25$ & $-0.77$ & & $-8.29$ & $-0.83$ & & $-14.34$ & $-0.97$ \\
$9.96$  &  & & $-6.33$ & $-0.75$ &  & & $-10.11$ & $-0.62$ & & $-7.03$ & $-0.57$ & & $-8.86$ & $-0.60$ \\
$14.94$ &  & & $-5.86$ & $-0.68$ &  & & $-9.65$  & $-0.55$ & & $-6.36$ & $-0.46$ & & $-6.13$  & $-0.44$ \\ \hline \hline 
\end{tabular}
\end{table*}

\textbf{Number of bound ions and effective charge} The cumulative number of ions of species $i$ as a function of the distance $r$ from the COM of dPGS is calculated as
\begin{equation}
N_{\mathrm{acc},i}(r) = \int^{r}_{0} c_i(r') 4\pi {r'}^2 \mathrm{d}r' \quad  \quad i = ++,+,-.
\label{nbi}
\end{equation}
Summing up the contribution of all charged species, the cumulative charge valency of the system as a function of the distance $r$ reads
\begin{equation}
Z_\mathrm{acc}(r) = \zd(r) + 2N_{\mathrm{acc},\submg}(r) + N_{\mathrm{acc},\subna}(r) - N_{\mathrm{acc},\subcl}(r),
\label{neff}
\end{equation}
where $\zd(r)$ denotes the spatial distribution of bare charge valency of the dPGS, obtained from the simulation.  
With that, the number of the bound ions and the effective charge valency of the dPGS follow as $\Nib = N_{\mathrm{acc},i}(\reff)$ and $\zeff = Z_{\mathrm{acc}}(\reff)$, respectively. The values are shown in Table~\ref{cg_table}. $\zeff$ and $\phieff$ exhibit a strong decrease in the magnitude with higher $\ctmg$, indicating an enhanced dPGS charge renormalization.

\section{Theoretical models}
\subsection{Basic model}\label{basic}
\begin{figure}[ht!] 
   \centering
      \includegraphics[width=5.5cm, height=5.5cm, keepaspectratio]{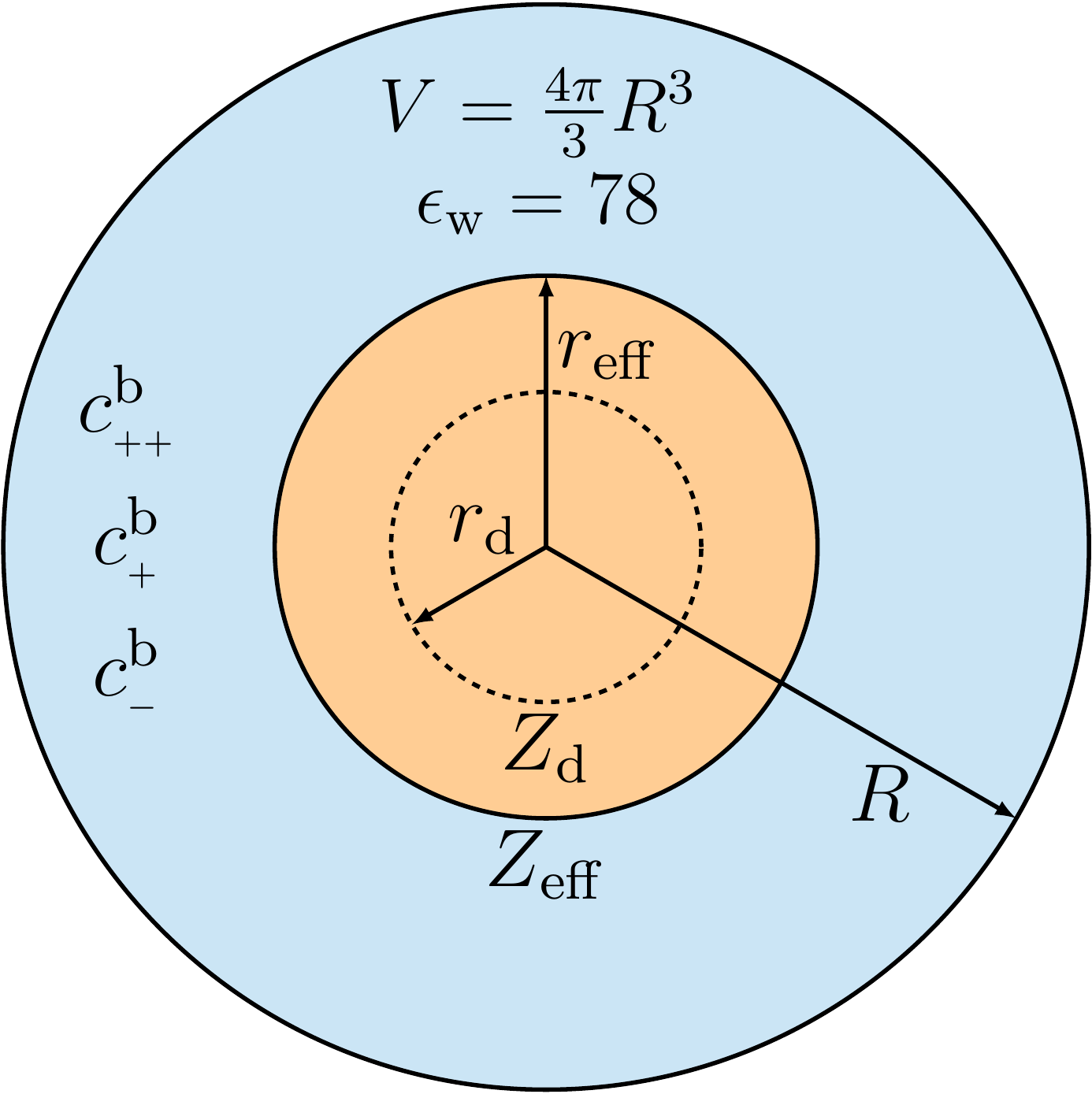}
   \caption{\label{fig:theoretical-model} Schematic of a theoretical model representing the system shown in Fig.~\ref{snap}(d). 
The computational cell domain (blue) is assumed to be spherical with the same volume as that of the simulation box, $V$, and with a uniform dielectric constant of water $\epsilon_\mathrm{w}=78$. 
dPGS is assumed to be a perfect sphere (orange) at the center of the domain. The dPGS bare and effective charge valencies are $\zd$ and $\zeff$, respectively.
$\rd$ is the  bare radius of dPGS, while
$\reff$, the effective radius, representing the distance separating the electric double layer regime ($r > \reff$) from the non-linear counterion 'condensation' regime ($r < \reff$).}
\end{figure}

In our theoretical models, the macromolecule is represented as a perfect sphere with the bare radius $\rd$, the bare charge valency $\zd$, 
the effective radius $\reff$ and the effective charge valency $\zeff$, enclosed in a spherical domain of radius $R$ and volume $V$, as shown in Fig.~\ref{fig:theoretical-model}. The total number of charged monomers in the macromolecule is $\Ns$, each of which is negatively charged with a charge valency $\zs$. 

All ionic species and the macromolecule are assumed to be in an aqueous bath with an implicitly modeled solvent, having a uniform dielectric constant $\epsilon_\mathrm{w}=78$ at a temperature $T=298$\,K. 

\subsection{\label{dm} The Donnan model (DM)}
The arguably simplest model for competitive uptake is the Donnan model. 
The Donnan equilibrium assumes two strictly electroneutral and mutually exclusive regions, \textit{i.e.}, the macromolecule region with the Donnan radius set to be the bare radius $\rd$ taken from simulations (\textit{i.e.}, with a bare macromolecular volume $\vd=4\pi \rd^3 /3$) and total homogeneously distributed bare charge of valency $\zd = \zs \Ns$ with a concentration $\cs = \Ns/\vd$ of charged groups of the macromolecule, and the bulk region outside the molecule with a bulk ion concentration $\ci$ ($i=+,+\!+,-$).  
Charge neutralization of the macromolecule by the counterions leads to the Donnan potential, which is a potential having a constant non-zero value in the macromolecule region. The potential in the bulk region is set to zero. 
The equilibrium distribution (partitioning) of ions among the regions results in the concentrations of ionic species $i$ as $\cbi$ and $\ci$ in the macromolecule and bulk regions, respectively.  These concentrations are related via the partition coefficient $\ki$, given by
\begin{equation}
\ki = \frac{\cbi}{\ci} \quad \quad i=+\!+,+,-
\label{ki-don1}
\end{equation}
Neglecting ion--ion correlations, an approximate expression for $\ki$ can be obtained using the condition that the equilibrium electrochemical potential of ion $i$ is equal in both the macromolecule and bulk regions, implying that
\begin{equation}
\lin\,\ci =  z_i\phid + \lin\,\cbi + \beta \dgi
\label{phase_eqm}
\end{equation}
where $\phid$ is the dimensionless Donnan potential (scaled by $\kt/e$) in the macromolecule region and $\beta^{-1} =\kt$ is the thermal energy.
With $\dgi$ we account for additional non-electrostatic effects that can drive adsorption, \textit{e.g.}, dispersion and hydrophobic forces in the net ion--macromolecule interaction, and is termed the ion-specific binding chemical potential of the condensed ion. The inclusion of $\dgi$ has been considered in previous work, for example, as a term reflecting the steric ion--ion packing effects in a Donnan model for ion binding by polyelectrolytes or charged hydrogels.~\cite{Chudoba2018,arturo2014,Ahualli2014}  

Eq.~\eqref{phase_eqm} with the help of Eq.~\eqref{ki-don1} then leads to 
\begin{equation}
\ki = \frac{\cbi}{\ci}  = \mathrm{e}^{-\beta \dGi}= \mathrm{e}^{-\beta \dgi} \mathrm{e}^{-z_i \phid}
\label{ki_don2}
\end{equation}
where $\dGi$ is the total transfer chemical potential for ion $i$ from the bulk to the macromolecule region. This allows us to define the intrinsic partition ratio for ionic species $i$ as
\begin{equation}
\kiz = \mathrm{e}^{-\beta \dgi} \quad \quad i=+,+\!+
\label{kiz}
\end{equation}
and the Donnan partition ratio as a contribution from pure electrostatic interaction between the ion and the macromolecule environment as 
\begin{equation}
\kiel = \mathrm{e}^{-z_i \phid} \quad \quad i=+,+\!+
\label{kid}
\end{equation}
The electrostatic component of total binding chemical potential of a counterion $i$ is then defined as $\beta \dgiel = -\lin\,\kiel = z_i\phid$. Eq.~\eqref{ki_don2} can then be conveniently shortened as
\begin{equation}
\ki = \kiel \, \kiz
\label{ki-don3}
\end{equation}
where $\ki$ is shown as a composition of intrinsic and electrostatic effects. 
The signature assumption behind the Donnan model is the electroneutrality in the macromolecule region expressed as
\begin{equation}
\zs\cs + \sum_i z_i \ci\,\ki  = 0   
\label{neutral}
\end{equation}
Solving Eq.~\eqref{neutral} for $\phid$ enables us to evaluate the net partition coefficient $\ki$. 
Eq.~\eqref{neutral} has no closed solution for multivalent ions, but it exists for the case of only monovalent ions in the system, ($i=\pm$) and is given as~\cite{cemil2}
\begin{equation}
\phid = -\lin \left(-\frac{ \sqrt{1 + \chi_{\subna} \chi_{\subcl}} + 1 }{\chi_{\subna}} \right)
\label{donnan-mono-expr}
\end{equation}
where $\chi^{}_i = 2\kiz \, \ci /\zs\cs $. Note that $\chi^{}_i<0$, since the valency of charged groups $\zs$ is negative. Using Eqs.~\eqref{ki_don2}, \eqref{ki-don3} and \eqref{donnan-mono-expr}, for the monovalent-only case, the number of ions of species $i(=\pm)$ partitioned into the macromolecule region is then given as 
\begin{equation}
N^\mathrm{b}_{\pm} = c^0_{\pm} \vd \, \mathcal{K}_{\mathrm{int}, \pm}  \left( -\frac{ \sqrt{1 + \chi_{\subna} \chi_{\subcl}} + 1 }{\chi_{\subna}} \,\right)^{\pm 1}  
\label{Nb-don}
\end{equation}
To evaluate the competition between MCs and DCs in the Donnan model we evaluate Eqs.~\eqref{ki_don2} and \eqref{neutral} numerically, \textit{cf.} section~\ref{numerical}.

Because of the electroneutrality assumption, the Donnan prediction for the amount of counterion sorption by the macromolecule in the monovalent-only case is given by $\Nab = |Z| + N^\mathrm{b}_{\subcl}$. For highly charged macromolecules, \textit{i.e.}, $\chi_i \rightarrow 0$, Eq.~\eqref{Nb-don} this trivially yields $\Nab \simeq |Z|$. For the competitive sorption case, however, the result is non-trivial and  can give a useful orientation with little effort. The Donnan model should become more quantitative for large dPGS generations, i.e., large size and/or high salt concentrations, so that $\kappa \rd \gg 1$,  for which the electroneutrality assumption is then very well justified.

\subsection{\label{sec:ppbl} Ion-specific Penetrable Poisson--Boltzmann (PPB) model}
We now put forward a penetrable PB (PPB) model in which the charge profiles can be resolved in $r$, the radial distance from the macromolecular center.  Since our charged macromolecules we have in mind (dPGS above and similar) are polymer-based with open structures and typically internally smeared out charge distributions, we opted (as in the Donnan model) for a penetrable model instead of a PB model for surface adsorption as typically used in studies of colloidal charge renormalization.~\cite{Wall1957,Ohshima1982,Ohshima2008} 
Based on the parametrization described in the basic model (Section~\ref{basic}), we assume the macromolecule as a perfect penetrable sphere with a charge valency $\zd = \zs \Ns$ and radius $\rd$, as shown in Fig.~\ref{fig:theoretical-model}. 
$\rd$ is taken from the dPGS internal charge distribution obtained from simulations, \textit{cf.} Section~\ref{sim_results} and Fig.~\ref{rho_sulfate}.
The charged monomers of the macromolecule, thus, have a uniform number distribution $\cs = \Ns/\vd$ (where $\vd = 4\pi \rd^3/3$) within the volume $\vd$.
$\cs$ is applicable only within the macromolecule domain, \textit{i.e.}, $\cs(r) = \cs\left(1-H(r-\rd)\right)$, where $H(r)$ is the Heaviside-step function. 
As an improvement to the standard PB model, here we also consider a contribution of the intrinsic non-electrostatic ion-specific interaction $\dgi$ between the ion and the macromolecule,~\cite{Kalcher2010,Kalcher2010a} analogous to Eq.~\eqref{phase_eqm} in the Donnan model above. 
Assuming the electrostatic potential far away from the macromolecule, $\phi\left(r \rightarrow R \right)=0$, we first balance the chemical potential for each ion, between the bulk regime far from the macromolecule and the regime at the finite distance $r$ from the center of the macromolecule
\begin{equation}
\lin\,\ci = z_i\phi(r) + \lin\,c_{i}(r) + \beta \dgi (r), 
\label{pb_chem_pot_bal}
\end{equation}
which is similar to Eq.~\eqref{phase_eqm}, but in a distance-resolved manner. $\dgi$ is considered on a local level, \textit{i.e.}, $\dgi(r) = \dgi\left(1-H(r-\rd)\right)$. The Boltzmann ansatz then becomes
\begin{equation}
c_i(r) = \ci \,\mathrm{e}^{-z_i\phi(r) - \beta \dgi(r) }
\label{pb_ansatz0}
\end{equation}
The distance-resolved electrostatic potential can be calculated from Eq.~\eqref{pb_ansatz0} together with the Poisson's equation as
\begin{equation}
\nabla^2 \phi(r) = -4\pi \lb \left(\sum_{i} z_i c_i(r) + \zs\cs(r) \right) \quad \quad i = +\!+,+,-
\label{pb-pot}
\end{equation}
which establishes the PPB model including ion-specific binding effects.  The boundary conditions used are $({\dd}\phi/{\dd}r)\left(r \rightarrow 0 \right) = 0$ and $({\dd}\phi/{\dd}r)\left(r \rightarrow R \right)= 0$.  

An effective radius for dPGS is calculated independently for this model (labeled $\reff^\mathrm{PB}$) using the Alexander prescription~\cite{alexander1984charge,Trizac2002,bocquet2002effective,Levin2004} on the obtained potential $\phi$, the same recipe used to calculate $\reff$ from simulations, \textit{cf.} Section~\ref{sim_analysis}. 
The values of $\reff^\mathrm{PB}$ for G$_2$-dPGS and G$_4$-dPGS are obtained as $1.42$\,nm and $2.36$\,nm, respectively, under the operated range of $\ctmg$ and at $\ctna=150.37$\,mM.
The $\reff^\mathrm{PB}$ values are thus found to be close to those obtained from the simulations, as shown in Table~\ref{cg_table}. 
The effective surface potential of the macromolecule is then given by $\phieff^\mathrm{PB} = \phi(\reff^\mathrm{PB})$.
The number of bound ions of species $i$ within $\reff$, is then given by
\begin{equation}
\Nib = \int_0^{\reff} c_i(r)\, 4 \pi r^2 \dd r \quad  \quad  i=+,+\!+
\label{Nib}
\end{equation}
The corresponding effective charge valency $\zeff^\mathrm{PB}$ is calculated using Eq.~\eqref{neff}. 
The PPB equations are solved numerically, \textit{cf.} Section~\ref{numerical}. 

\subsection{\label{mvh} Manning--McGhee--von~Hippel binding model (MMvH)}
In this section, we introduce a model based on a discrete two-state (condensed or free) perspective for the counterions, built to capture the essential physics of polyelectrolyte--ion binding in an accurate but minimalistic fashion. The model is an extension of ideas by Manning,~\cite{Manning2007} in which ion-condensation on charged spherical surfaces was described on a mean-field free energy level as a competition between the charging (Born) self-energy of the macromolecule  in salt solution and the entropy cost of binding for one-component counterions.  Here, we extend this model to the case of mixtures of MCs and DCs, including binding saturation for a fixed number of binding sites like in Langmuir isotherms. The extension of the latter to binary binding of one or two binding sites by mono- or divalent solutes, respectively, was put forward buy McGhee and von Hippel.~\cite{Mcghee1974} Therefore, we name the model Manning--McGhee--von Hippel binding model (MMvH).
 
Following Manning,~\cite{Manning2007} we treat the macromolecule as an impenetrable sphere of radius $\reff$ and charge valency $\zd = z_s\Ns$ taken from simulations, and extend the Manning's model into a discrete binding site model, where the $\Ns$ charged monomers act as a finite collection of discrete binding sites for both the MCs and DCs.  For the case of the DCs, two adjacent charged monomers can collectively act as a single binding site for a DC.  The resulting combinatorial ways to arrange the bound MCs and DCs lead to mixing entropies worked out by McGhee and von Hippel.~\cite{Mcghee1974} 
Pertaining to the canonical ensemble, we fix the total number of salt ions $n_i$, the corresponding concentrations $\cti$ ($i= +\!+,+,-$), the number of monovalent counterions $\Ns$ to the macromolecule, the total number of binding sites on the macromolecule and the total domain volume $V$.  The coions in this model simply serve the function of maintaining electroneutrality in the total domain and their explicit adsorption is neglected. 
  
A counterion $i$ ($=+,+\!+$) is assumed to bind to the macromolecule and to occupy $f_i$ consecutive (spatially adjacent) charged terminal groups of the macromolecule.   We designate $f_{\subna}=1$ and $f_{\submg}=2$ for MCs and DCs, respectively, implying that, in a bound state, one MC occupies only one charged terminal group, while one DC occupies two consecutive charged terminal groups, owing to the fact that each terminal group has a charge valency $\zs = -1$. 
Consider at a given state, $\Nab$ MCs and $\Nbb$ DCs are bound to the macromolecule. 
The \textit{binding density}, \textit{i.e.}, the number of bound counterions per charged terminal group is then $\Nab/\Ns$ and $\Nbb/\Ns$ for MCs and DCs, respectively. 
By multiplying with $f_i$, we then define the fraction of the binding sites occupied by the counterions, \textit{i.e.}, coverages $\ta = f_{\subna} \Nab / \Ns = \Nab / \Ns $ and $\tb = f_{\submg} \Nbb/\Ns= 2\Nbb/\Ns$.  
Thus, the total number of binding sites on the macromolecule available for MCs, is $N_{\subna} = \Ns/f_{\subna} = \Ns$, and those available for DCs, is $N_{\submg} = \Ns/f_{\submg} = \Ns/2$. 
The effective charge valency of the macromolecule is then $\zeff = -\Ns + \Nab + 2\Nbb = {-\Ns(1-\ta-\tb)}$. The total Helmholtz free energy $\ft$ depends on the coverages $\ta$ and $\tb$ and the ionic concentrations $\cti$. The coverages can then be obtained by minimizing $\ft$ simultaneously with respect to $\ta$ and $\tb$. 
The total Helmholtz free energy $\ft$ is given by the expression
\begin{equation}
\ft = \fel + \fid + \fmx + \fin
\label{eq:mvh_ftot}
\end{equation}
where the four additive contributions, $\fel$, $\fid$, $\fmx$ and $\fin$ are defined respectively as (i) electrostatic (Born) self-energy of charge renormalized macromolecule, (ii) ideal gas entropy of free ions in the bulk regime, (iii) mixing entropy of the condensed counterions in the macromolecule, and (iv) the non-electrostatic ion-specific binding free energy between the condensed counterion and the corresponding binding site on the macromolecule.

The Born charging self-energy of the macromolecule immersed in an electrolyte solution associated with the Debye screening length $\kappa^{-1}$, refers to the work required to charge the macromolecule from its electroneutral to a certain charged state. 
Following Manning, such a charged state is associated with the effective charge $\zeff e$, corresponding to the sum of the intrinsic bare charge of the macromolecule $\zd$ and its captive, neutralizing counterions.~\cite{Manning2007} 
Thus, the expression for the Born charging free energy of the macromolecule (or the self energy of the charge renormalized macromolecule) per monovalent binding site is thus expressed as 
\begin{equation}
\beta \fel = \frac{\zeff^2 \lb}{2 \Ns \reff(1+\kappa \reff)} = \frac{\zeta}{2} (1-\ta-\tb)^2
\label{fel1}
\end{equation} 
where $\zeta/2$ is the Born free energy per monovalent binding site in the absence of counterion condensation, and $\zeta$ is given for surface charging by\cite{mcquarrie2000statistical}
\begin{equation}
\zeta =  \frac{\Ns \lb }{\reff(1 + \kappa\reff)}
\label{zeta}
\end{equation}

Considering the effective volume of dPGS $\veff$ to be very small compared to the total volume $V$ ($\veff \ll V$), the bulk concentrations of MCs and DCs are given by
\begin{align}
\begin{split}
&\cna = \ctna + \frac{\Na(1-\ta)}{V} \\
&\cmg = \ctmg - \frac{\Nb\tb}{V} 
\end{split}
\label{eq:bulk_conc_mvh}
\end{align}
owing to the depletion of the ions in the bulk due to partitioning. 
$\cna$ above is calculated considering the monovalent counterions remaining in the solution, in the salt-free limit.
We assume that no anions are bound to the macromolecule binding sites, hence their bulk concentration is assumed to be the same as their salt concentration, \textit{i.e.}, $\ccl = \ctcl$.

The ideal gas free energy of free cations in the bulk, normalized by the number of monovalent binding sites $\Ns$, is given as\\
\begin{align}
\begin{split}
&\beta \fid = -\frac{S_\mathrm{id}}{\Ns \kB} = \sum_{i=+,++} \left(\frac{n_i - \Nib}{\Ns}\right)\left(\lin\,\ci\Lambda^3_i -1\right) \\
&= \sum_{i=+,++} \left(\frac{n_i - N_i \ti}{\Ns}\right)\left[\lin \left(\cti \Lambda^3_i - \frac{N_i \ti \Lambda_i^3}{V}\right) -1\right]
\end{split}
\end{align}
where $\Lambda_i$ and $n_i$ are the thermal (de Broglie) wavelength and the number of salt ions $i$. 

The bound DCs and MCs can occupy the binding sites on the macromolecule in different proportions, and can distribute among the occupied sites in multiple ways at a certain bound coverages $\ta$ and $\tb$. We exert constraints to such possibilities of binding compositions and configurations, such that, ($i$) one bound DC can only bind to two adjacent monovalent binding sites, ($ii$) all non-overlapping configurations between the bound ions are possible, ($iii$) there are no designated binding sites for DCs, and ($iv$) the position of the bound DC can be shifted by a single adjacent monovalent binding site. The number of possible combinatorial binding arrangements under these constraints, adopted from the work by McGhee and von Hippel,~\cite{Mcghee1974} is given by 
\begin{equation}
W =  \frac{ \gamma^{\Nab}_{\subna} \gamma^{\Nbb}_{\submg} (\Ns - \Nbb)!}{\Nab!\Nbb!(\Ns - 2\Nbb - \Nab)!}
\label{w}
\end{equation}
where we define $\gamma_i=\vi/\Lambda^3_i$ in terms of the effective configurational volume $\vi$ in the bound state.~\cite{XiaoCPS} 
$\vi$ takes into account the rotational and vibrational degrees of freedom of a bound counterion $i$.
We now define the free energy associated with the partition function $W$, normalized by the number of monovalent binding sites $\Ns$, as the free energy of mixing of the bound ions per binding site,
\begin{align}
\begin{split}
&\beta \fmx = -\frac{S_\mathrm{mix}}{\Ns k_{\rm B}}= -\frac{1}{\Ns}\lin \, W\\
 & \simeq \ta\lin \ta + \frac{\tb}{2}\lin \frac{\tb}{2} -\left(1-\frac{\tb}{2}\right)\lin \left(1-\frac{\tb}{2}\right) \\
&+  (1 - \ta - \tb)\lin (1 - \ta - \tb) \\
&- \ta \lin \frac{\va}{\Lambda_+^3} - \frac{\tb}{2} \lin \frac{\vb}{\Lambda_{++}^3}
\end{split}
\label{fmx}
\end{align}
where the Stirling approximation has been used for the logarithm of the factorials.  This description of condensed counterion entropy is different than the ion-binding models proposed in previous works for linear polyelectrolytes~\cite{flory1953molecular,Raphael1990,Muthukumar2004} in terms of the localization of counterions within volume $\vi$. 

We express this intrinsic interaction $\fin$  by the intrinsic binding chemical potential $\dgi$ of each bound ion $i$. The sum of such interactions for all bound ions, normalized by the total number of monovalent binding sites gives
\begin{align}
\begin{split}
\beta \fin &= \frac{1}{\Ns}\left(\Nab \beta \dga + \Nbb \beta \dgb\right) \\
&= \ta \beta \dga + \frac{\tb}{2}\beta \dgb
\end{split}
\end{align}

The equilibrium coverages $\Theta_i$ are then obtained by the minimization condition
\begin{equation}
\frac{\partial}{\partial \ti} \ft \overset{!}{=} 0 \qquad \quad i=+,+\!+
\end{equation}
This leads to the relation
\begin{equation}
\dgicr + \dgiel + \dgimx + \dgi = 0 \quad  \quad i=+,+\!+
\label{k_and_g_all}
\end{equation}
where $\dgicr$ denotes the translational entropy change associated with one ion $i$ when it transfers from the bulk environment to the bound state in the macromolecule. 
$\dgiel$ is the \textit{electrostatic binding chemical potential} and $\dgimx$ is the \textit{mixing chemical potential}. 
Eq.~\eqref{k_and_g_all}, similar to the PPB (Eq.~\eqref{pb_chem_pot_bal}) and DM (Eq.~\eqref{phase_eqm}) models, indicates the counterion chemical potential components contributing to its condensation on the macromolecule.
The expressions for the constituent chemical potential contributions in Eq.~\eqref{k_and_g_all} are given by
\begin{align}
\begin{split}
&\beta \dgicr =-\lin\,\ci\vi  \qquad \qquad \qquad  \quad i=+,+\!+\\
&\beta \dgiel =-z_i\zeta(1-\ta-\tb) \quad  \quad i=+,+\!+\\
&\beta \dgimx =
\begin{cases}
\begin{aligned}
 \lin \frac{\tb \left(2-\tb \right)}{4(1-\ta-\tb)^2} \quad \quad i=++\\[1.1ex]
 \lin \frac{\ta}{(1-\ta-\tb)} \quad  \quad i=+
\end{aligned}
\end{cases} 
\end{split}
\label{all_free_energies}
\end{align}

Using Eqs.~\eqref{k_and_g_all} and \eqref{all_free_energies} leads to the final form of the MMvH model, given by
\begin{equation}
K_{\submg} = v^0_{\submg} \kbz \mathrm{e}^{2\zeta \left( 1-\ta - \tb \right)} = \frac{\tb (2 - \tb)}{4 \cmg {(1 - \ta - \tb)}^2} 
\label{Kb}
\end{equation}
\begin{equation}
K_{\subna} = v^0_{\subna} \kaz \mathrm{e}^{\zeta \left(1-\ta - \tb \right)} = \frac{ \ta }{\cna (1 - \ta - \tb)} 
\label{Ka}
\end{equation}
where $K_i$ are the \textit{equilibrium binding constant} associated with the binding of ion $i$ to its corresponding binding site on the macromolecule. 
The relationship between $K_i$, the total binding chemical potential $\dGi$ and the total partition ratio $\ki$ is given as 
\begin{equation}
\beta \dGi = -\lin\,\frac{K_i}{\vi} = -\lin\,\ki \quad  \quad i=+,+\!+
\label{kbind}
\end{equation}
Or in other words, referring back to Eq.~\eqref{ki-don3},
\begin{equation}
\ki = \kiz \,\kiel= \kiz \, \mathrm{e}^{z_i \zeta \left(1-\ta - \tb \right)}
\end{equation}
where the electrostatic contribution of the total partition ratio is defined as
\begin{equation}
\kiel = \mathrm{e}^{-\beta \dgiel} = \mathrm{e}^{z_i\zeta(1-\ta-\tb)} \quad  \quad i=+,+\!+
\label{kiel-mvh}
\end{equation}
From Eq.~\eqref{kbind}, for a given magnitude of $K_i$, the absolute magnitude of $\dGi$ depends on $\vi$, which we calculate from our simulations and predict respective values of $\dGi$.

Finally, we consider the limit of the MMvH model for vanishing DCs (MCs only). Without DCs, we have 
\begin{align}
\begin{split}
&\beta \dgcr =-\lin\,\cna v^0_{\subna}  \\
&\beta \dgel =-\zeta(1-\ta) \\
&\beta \dgmx = \lin\,\frac{\ta}{(1 - \ta )}
\label{g_langmuir}
\end{split}
\end{align}
Combining Eqs.~\eqref{k_and_g_all} and \eqref{g_langmuir} leads to the ``Manning--Langmuir" (ML) model
\begin{equation}
K_{\subna} = v^0_{\subna} \kaz \mathrm{e}^{\zeta (1 - \ta )} = \frac{\ta}{\cna (1 - \ta )} 
\label{manning-langmuir}
\end{equation} 
The McGhee--von Hippel combinatorics here reduces to the standard one-component Langmuir picture, \textit{i.e.}, the right-hand-side of Eq.~\eqref{manning-langmuir} reflects the Langmuir isotherm. The standard Langmuir model is thus extended to include charging free energies by ion condensation (charge renormalization) and ion-specific binding. From another perspective, it extends the Manning model for the counterion condensation on spheres~\cite{Manning2007,Gillespie2014} to include ion-specific effects as well as the saturation of binding sites in terms of the translation entropy of the condensed ions.

\subsection{Numerical evaluation} \label{numerical}
The PPB model, with the assumption of the uniform intrinsic macromolecular volume charge distribution $\cs(r)e$ and with the knowledge of the bare radius $\rd$ of the macromolecule inherited from simulations, generates the distance-resolved number density profiles of charged species, similar to Fig.~\ref{density_ions}. 
Hence, it performs the same analysis as that for simulations (\textit{cf.} Section~\ref{sim_analysis}), to calculate the effective radius $\reff$ and other electrostatic properties of the macromolecule, such as $\zeff$, $\phieff$, etc.
The DM model also assumes uniform $\cs(r)e$ and requires the knowledge of the electroneutrality radius, which is taken as $\rd$ from simulations as an input parameter, similar to the PPB model.
The MMvH (ML) model, on the other hand, assumes the macromolecule as a hard sphere with a uniform surface charge distribution. 
The effective radius of the hard sphere $\reff$ is taken from simulations as an input parameter. 
The results from the DM, PPB and MMvH (ML) models and simulations are compared in terms of the coverages $\ti$ ($i=+\!+, +$), which are defined as $\ti =\Nib/\Ni$, where $\Nib$ is the number of condensed counterions $i$ and $\Ni$ is the corresponding number of binding sites available on dPGS, defined in Section~\ref{mvh}.
Since the PPB model deals with a volume sorption, while the DM model deals with the ion partitioning between two electroneutral phases, ``coverage" $\ti$ in these cases are interpreted as a load or an extent of neutralization of dPGS.
For the DM, PPB and MMvH (ML) models, the intrinsic partition coefficients $\kiz$ for both ions ($i=+\!+,+$) are unknowns and taken as fitting parameters in order to match the coverages from the simulations, which are described  in Section~\ref{sec:cg_ff}. 
Regarding the PPB and DM models, we make a further assumption that intrinsic non-electrostatic ion--binding site interaction for the MCs is identical to that for the monovalent anions, \textit{i.e.}, $\kaz = \kcz$. 

Mathematically, the PPB model represents a boundary-value problem having a second order differential equation (Eq.~\eqref{pb-pot}) non-linear in the electrostatic potential paired with the boundary conditions, while the MMvH model (Eqs.~\eqref{Kb} and \eqref{Ka}) represents two non-linear simultaneous equations in coverages $\ta$ and $\tb$. 
Both PPB and MMvH models are evaluated self consistently for the potential and coverages, respectively. To solve Eq.~\eqref{pb-pot}, we employ the \texttt{solve\_bvp} function in the SciPy library (version 1.3.1) from Python (version 3.7.4), which solves a boundary-value problem for a system of ordinary differential equations using the fourth order collocation algorithm.~\cite{kierzenka2001bvp} The bulk concentration $\ci$ is obtained using the law of conservation of mass, in an iterative manner. 
Eqs.~\eqref{Kb} and \eqref{Ka} are solved using \texttt{fsolve} function from the SciPy library, which is also used to evaluate the DM model (Eq.~\eqref{neutral}) representing the single non-linear equation in the Donnan potential $\phid$.

The effective configurational volume $\vi$ of bound counterions, used in the MMvH model is assumed to be equal for both counterions, \textit{i.e.}, $v_{\submg} = v_{\subna} = v_0$. 
It is worth considering that the volume $v_0$ depends on the precise nature of the bound state and it is infeasible to have its knowledge in experiments due to unknown microscopic details, although it can be computed using simulations.~\cite{xu:biomacro,Yu2015}
According to the convention in experiments, the standard volume is defined as $v_0 = 1\,\mathrm{M}^{-1} \simeq 1.6$\,nm$^{3}$, corresponding to the standard concentration $c^\mathrm{std} = 1$\,M.~\cite{atkins,Gilson2007,General2010a} 
In this case, the total binding chemical potential $\dGi$ can be referred to as the standard binding energy $\Delta G^0$.~\cite{Gilson2007,General2010a}

\section{Results and Discussion}
\subsection{Monovalent limit: theoretical comparison and best fit to simulations}
\begin{figure*}[htbp!]
\centering
\includegraphics[width=12cm, height=12cm, keepaspectratio]{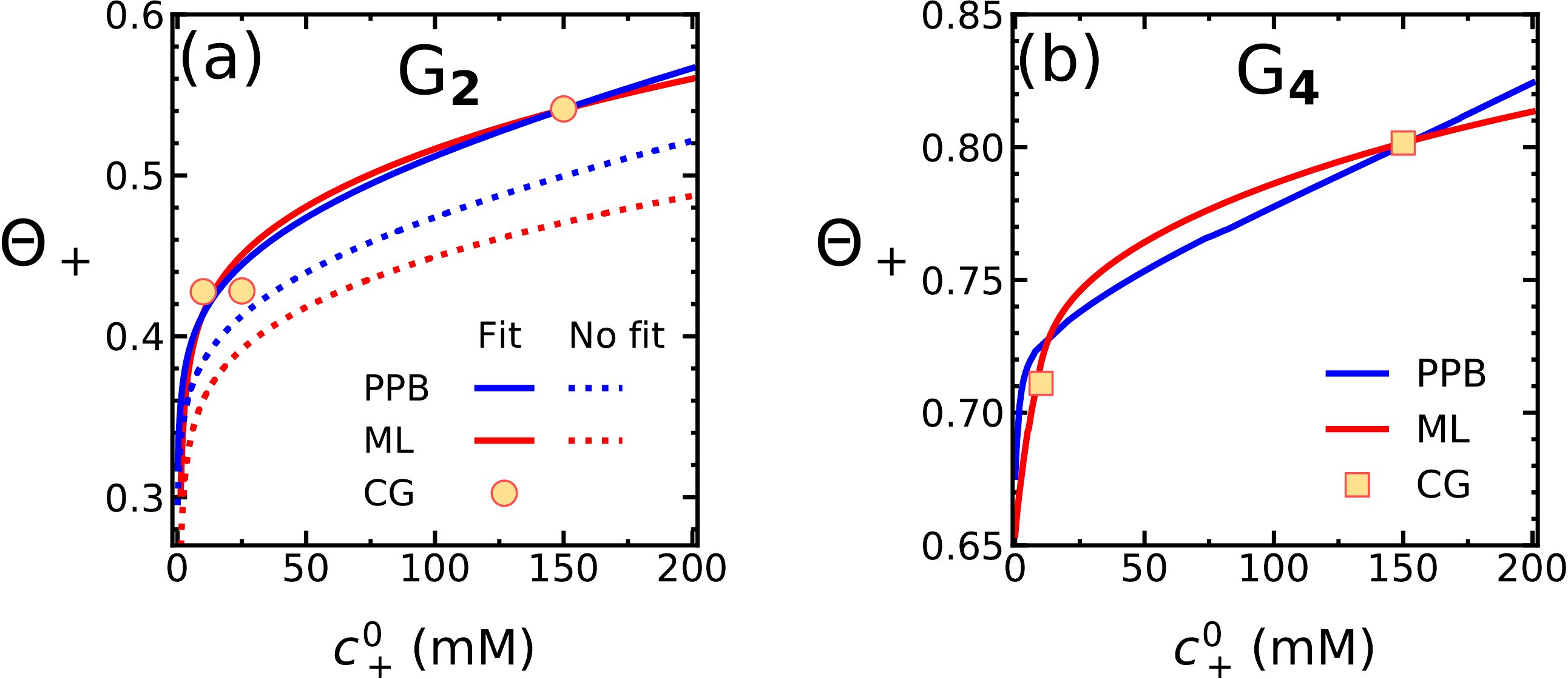}
\caption{\label{monocg} Model predictions [PPB (Eq.~\eqref{pb-pot}) and ML (Eq.~\eqref{manning-langmuir})] of the coverage $\Theta_{+}$ of MCs in the monovalent limit, as a function of the MC concentration $c^0_{+}$, compared with the CG simulations (circle and square symbols; CG). 
(a) For the case of G$_2$-dPGS, the dotted lines represent the results for vanishing intrinsic binding chemical potential $\dg{+}$, while the solid lines show the results obtained by fitting $\dg{+}$ to the simulations (yellow circles). 
The fitted $\dg{+}$ values obtained for the PPB and ML models are $-0.45\,\kt$ and $-1.81\,\kt$, respectively. 
The ML model uses the configurational volume $v_0=1.04$~M$^{-1}$ as obtained from our previous CG simulations.~\cite{xu:biomacro} 
(b) Comparison of binding coverages obtained by ML and PPB models for G$_2$-dPGS and G$_4$-dPGS. 
The dashed lines denote the model results fitted to G$_4$-dPGS simulations (yellow squares). The fitted values of $\dg{+}$ obtained for PPB and ML models are $-0.56\,\kt$ and $-1.85\,\kt$, respectively, fairly close to those obtained for G$_2$-dPGS. 
The configurational volume of bound ions for G$_4$ is fixed to $v_0=0.57$~M$^{-1}$ and is obtained from our previous simulations.~\cite{xu:biomacro} } 
\end{figure*}

Considering the monovalent limit as reference case, we now start with the application of aforementioned theoretical binding models.  
Fig.~\ref{monocg}(a) shows the predictions of the PPB and ML (monovalent-only limit of MMvH) models for the variation of the binding coverage of MCs, $\ta$, as a function of the MC concentration, $\ctna$.   
It can be observed that $\ta$ increases sharply for a small increase in $\ctna$ from $0$ to $\sim 10$\,mM, while it increases gradually for larger $\ctna$. 
This is attributed to the combined contribution of the electrostatics and an entropy of a bound counterion, facilitating condensation.  
In the low $\ctna$ regime, the bare charge of G$_2$-dPGS is weakly renormalized, and some of the dPGS binding sites are unoccupied. This leaves a high propensity of condensation for new incoming counterions.  This can be conveniently explained via the ML model. 
Referring to Eq.~\eqref{manning-langmuir}, the increase in the condensation of MCs at the limit of low $\ctna$, $\lim_{\ctna\to 0} {\dd}\ta/{\dd}\ctna$ 
is directly proportional to the total binding constant $K_{\subna}$, while at high $\ctna$, $\lim_{\ctna\to \infty} {\dd}\ta/{\dd}\ctna = 0$. This implies that at low $\ctna$, the resultant low coverage $\ta$ leads to a high electrostatic driving force for condensation as well as entropy of a bound counterion, thus a high amount of condensation. On the other hand, at high $\ctna$, the macromolecule charge is almost entirely renormalized and most of the binding sites are occupied, resulting in hardly any increase in condensation.

Comparing the coverage profiles from PPB and ML models that neglect ion-specific effects, \textit{i.e.}, with $\dga=0$ (dotted curves), we find that the PPB coverage values are close to the ML values in the low $\ctna$ regime, however, attain higher values than the ML counterpart at high $\ctna$. 
This is attributed to the effects of discrete binding sites incorporated in the ML model, in the form of the configurational volume $v_0$ (here, we used $v_0=1.04$~M$^{-1}$ obtained from our previous simulations~\cite{xu:biomacro}). 
The PPB model, on the other hand, assumes the condensed ions as point charges, leaving no entropic penalty for new incoming counterions as they condense on the binding sites.  
Another reason is that the PPB model also incorporates, to some extent, the non-linear effects in the electrostatic interactions, which are not considered in the DH-level Born energy used in the ML model.
Both models, however, underestimate the simulations if we do not include corrections via $\dga$. The reason is likely the approximative treatments of the electrostatic energy in both models, PPB and ML, which are mean-field and do not include the discrete nature of the charged binding sites and the complex spatial charge correlations inside the macromolecule. 
The DM model, in addition to these assumptions, takes the macroscopic view of macromolecule and bulk phases in a segregated form. 
The model then predicts the ion partitioning while imposing electroneutralities of phases.
In that respect, for highly charged macromolecules like dPGS, the DM model predicts $\Nab \simeq \Ns$, implying $\ta \simeq 1$. 
This plot is not shown, since it does not provide a useful insight for us in the context of counterion condensation.  
The case of salt concentration $\ctna = 0$ is referred to as the counterion-only case, and gives $\ta \sim 0.28$ for the PPB model. 
Note that $\ta$ in this limit is system specific, since the size of the simulation box/computational domain determines the counterion concentration and subsequently the coverage. 
The coverage $\ta$ in the ML model in this limit is undefined, since the electrostatic binding energy of MCs depends on the screening length $\kappa^{-1}$, which is undefined in this model in the absence of the salt. 

In the next step, $\ta$ values for PPB and ML models are fitted (bold curves in Fig.~\ref{monocg}(a)) to the simulation results for G$_2$-dPGS in the monovalent limit by allowing ion-specific effects in the counterion--macromolecule binding, \textit{i.e.}, $\dg{+}$ as a fitting parameter. 
The values of $\dg{+}$ are found to be $-0.45\,\kt$ and $-1.81\,\kt$ for PPB and ML models, respectively. 
Recall that the simulations have not really included ion-specific effects in terms of specific hydration phenomena, etc., still, they include excluded-volume, dispersion attraction, and importantly, all electrostatic charge--charge correlations, not captured in the mean-field theories. Hence, the ion-specific fitting parameters can be viewed in general as correction factors, including all ionic contributions that are beyond the mean-field treatment of the PPB and ML models. The larger fitting parameter for ML than PPB (in the absolute value) may indicate the higher level of approximations in the ML model. 
Having the models now informed using the benchmark data from simulations, they can be utilized to predict the binding at other ion concentrations.

Fig.~\ref{monocg}(b) shows the numerical fitting of $\ta$ values (dashed curves) to those obtained from G$_4$-dPGS simulations. 
The values of $\dg{+}$ as a fitting parameter are $-0.56\,\kt$ and $-1.85\,\kt$ for PPB and ML models, respectively, which are close to those obtained for G$_2$-dPGS, within the error difference of $\sim 0.1\,\kt$. 
The ML model fits better to both G$_2$-dPGS and G$_4$-dPGS CG results than the PPB model at large $\ctna$, which may indicate that the dPGS charge in the simulations acts more as finite binding sites, as assumed in the ML model.  

\subsection{Divalent case: theoretical comparison and best fit to simulations}
\begin{figure*}[htbp!]
\centering
\includegraphics[width=14cm, height=14cm, keepaspectratio]{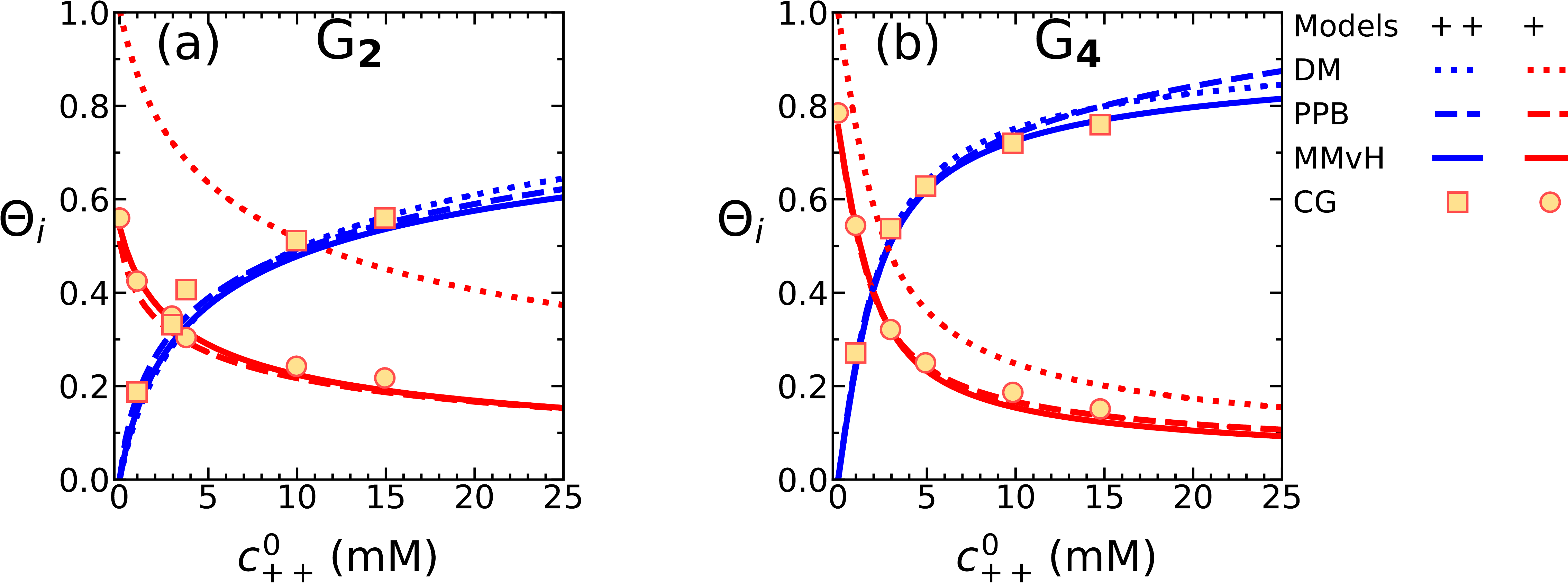}
\caption{\label{divalent} Coverages $\Theta_{+}$ and $\Theta_{++}$ on G$_2$-dPGS and G$_4$-dPGS obtained from the application of all models (MMvH, PPB, DM) as a function of the DC concentration $c^0_{++}$ in a mixture of DCs and MCs, compared to the CG simulations (yellow symbols; CG).  
The MC concentration, $c^0_{+}=150.37$~mM. 
Model $\Theta_i$ are fitted to simulations using the intrinsic binding chemical potentials $\dg{i}$ as fitting parameters. The values of $\dg{i}$ are obtained to be $-2.73\,\kt$ (G$_2$-dPGS) and $-2.98\,\kt$ (G$_4$-dPGS) for MMvH model, whereas $-1.77\,\kt$ (G$_2$-dPGS) and $-1.98\,\kt$ (G$_4$-dPGS) for PPB model.
The effective configurational volumes $v_0$ used in the MMvH model are $1.04$~M$^{-1}$ and $0.57$~M$^{-1}$ for G$_2$-dPGS and G$_4$-dPGS, respectively, and are obtained from our previous simulations.~\cite{xu:biomacro}
}
\end{figure*}
 
We now aspire to use the obtained $\dga$ to inform the MMvH and PPB models with the help of the reference data obtained from simulations, in order to capture the competitive ion binding in a mixture of MCs and DCs. The models fitted to the benchmark data can then be used to predict the binding coverages $\tb$ and $\ta$ for different dPGS generations and salt concentrations. In practice, we perform the numerical fitting of $\tb$ and $\ta$ obtained from the MMvH and PPB models to those from simulations, by fixing $\dga$ for MCs obtained from the monovalent-only case, and then subsequently fitting $\dgb$ for DCs. 
The values of $\dga$ for MCs obtained from the monovalent limit are, for a given binding model (ML or PPB), found to be approximately independent of the dPGS generation (with $\sim 0.1\,\kt$ as margin of error).
Therefore, $\dga$ is averaged over generations (G$_2$ and G$_4$), as shown in Table~\ref{mvh_ppb_kint_table}.
Fig.~\ref{divalent} depicts the behavior of MMvH, PPB and the DM model in terms of the binding coverages $\ti$, in a mixture of DCs and MCs. 
The MMvH model uses the effective configurational volumes $v_0=1.04$~M$^{-1}$ and $0.57$~M$^{-1}$ for G$_2$-dPGS and G$_4$-dPGS, respectively, as obtained from our previous simulations.~\cite{xu:biomacro} 
At low DC concentration, \textit{i.e.} in the monovalent limit, MCs act as the only counterions to the macromolecule, resulting in the highest MC coverage $\ta$. 
In this limit at $\ctna = 150.37$\,mM, both MMvH and PPB models show $\ta \simeq 0.57$ for G$_2$-dPGS, and $\ta \simeq 0.8$ for G$_4$-dPGS. 
As $\ctmg$ increases, more DCs bind to the macromolecule and more of the previously bound MCs get released into the bulk. 
Table~\ref{cg_table} shows the resultant effective charge valency $\zeff^\mathrm{PB}$ and potential $\phieff^\mathrm{PB}$ of G$_2$-dPGS and G$_4$-dPGS evaluated by the PPB model. 
Quantitatively consistent with the $\zeff$ and $\phieff$ obtained from simulations, $\zeff^\mathrm{PB}$ and $\phieff^\mathrm{PB}$ show a strong decrease in magnitude with a higher $\ctmg$, depicting higher dPGS charge renormalization.

\begin{table}[htbp!]
\caption{\label{mvh_ppb_kint_table} The values of the intrinsic component of the binding chemical potential $\dg{i}$ ($i=+,+\!+$) for the dPGS counterions for the Donnan (DM), PPB and MMvH models, obtained by the simultaneous numerical fit of the CG simulation coverages $\Theta_{+}$ and $\Theta_{++}$ to those obtained from the models (See Fig.~\ref{divalent}). 
The $\dg{i}$ values for a particular counterion species are averaged over G$_2$ and G$_4$ dPGS generations.
The MMvH model results are calculated for the configurational volume of a counterion in the bound state $v_0$ obtained from simulations and for $v_0=1$\,M$^{-1}$, which is the standard value typically considered in experimental evaluations of the standard binding energy.~\cite{atkins} 
The values of $v_0$ obtained from the simulations are $1.04$~M$^{-1}$ and $0.57$~M$^{-1}$ for G$_2$ and G$_4$-dPGS, respectively.~\cite{xu:biomacro} }
\setlength{\tabcolsep}{5pt} 
\renewcommand{\arraystretch}{1.3}
\begin{tabular}{cdd} \hline \hline
\text{Model} &  \multicolumn{2}{c}{$\dg{i}$ \text{($\kt$)}} \\ 
                                   &   ++   &   +   \\ \hline
\text{DM}                          &  5.13  &  3.37 \\
\text{PPB}                         & -1.87  & -0.50 \\ 
\text{MMvH} \:($v_0$ {\:\rm CG})   & -2.85  & -1.83 \\  
\text{MMvH} \:($v_0$ {\:\rm Std.}) & -2.86  & -1.44 \\ \hline \hline
\end{tabular}
\end{table}

\begin{figure*}[htbp!]
\centering
\includegraphics[width=11.8cm, height=11.8cm, keepaspectratio]{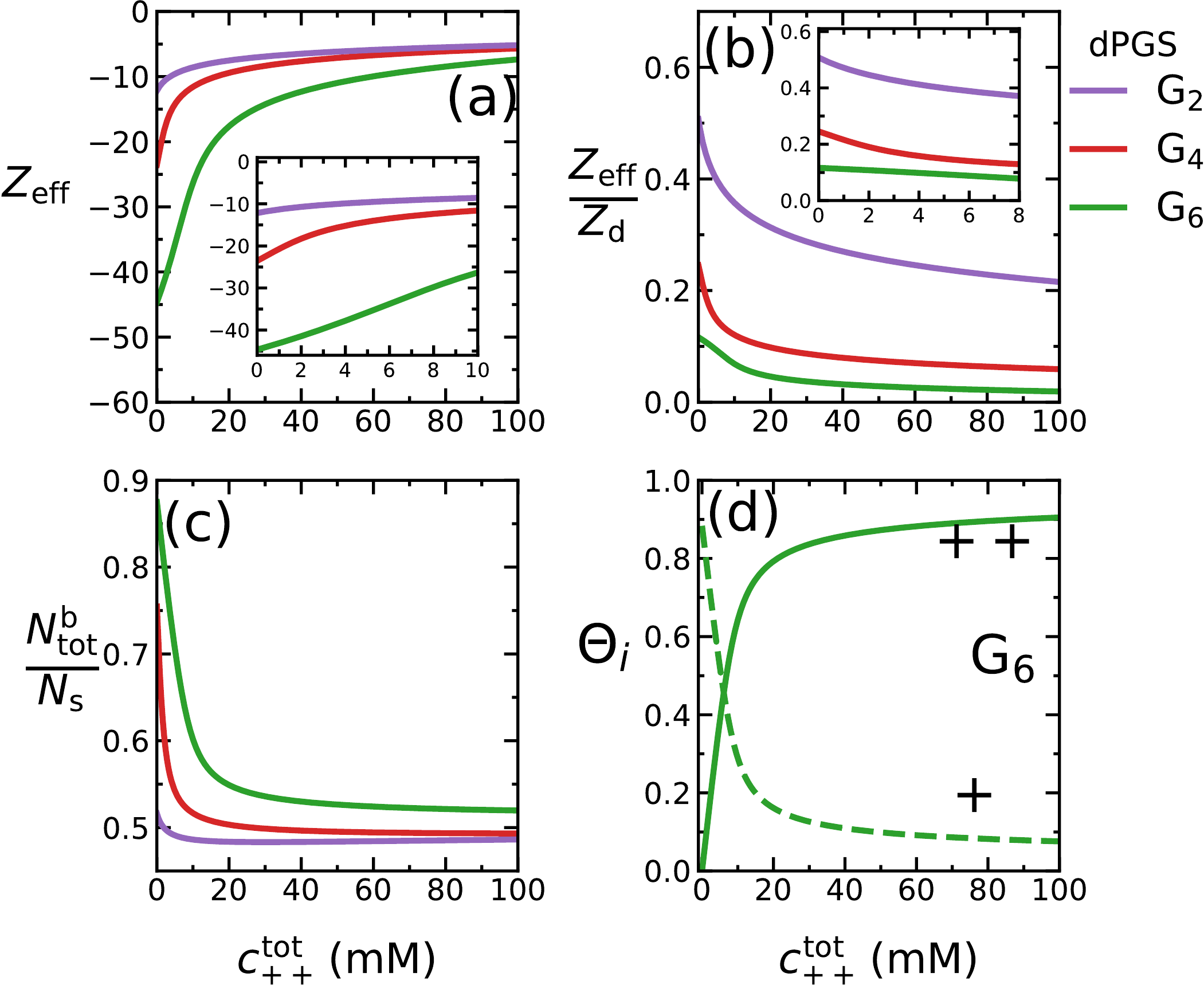}
\caption{\label{prediction} MMvH model predictions in a mixture of DCs and MCs.  
(a) The variation in the effective charge of dPGS with dPGS generation as a function of $c^0_{++}$.
(b) The effect of the dPGS generation on the ratio of the effective charge to bare charge $\zeff/\zd$ of dPGS as a function of $c^0_{++}$. 
The inset shows the smaller range of DCs concentrations, close to the physiological concentration range for DCs (Ca$^{2+}$ and Mg$^{2+}$ cations). 
(c) The variation in the total number of condensed counterions $N^\mathrm{b}_\mathrm{tot} = N^\mathrm{b}_{+} + N^\mathrm{b}_{++}$, normalized by the total number of MC binding sites $\Ns$, plotted as a function of $c^0_{++}$ for different dPGS generations. 
(d) Predicted values of binding coverages $\Theta_{+}$ and $\Theta_{++}$ for MCs vs. DCs competitive binding on G$_6$-dPGS.
The MC concentration $c^0_{+}$ is fixed to $150.37$\,mM. 
The intrinsic binding chemical potentials for DCs and MCs are fixed to the values averaged over generations (G$_2$ and G$_4$), $\dg{++}=-2.85\,\kt$ and $\dg{+}=-1.83\,\kt$, which are taken from simultaneous fitting of both coverages $\Theta_i$ ($i=+,+\!+$) to simulations (See Table~\ref{mvh_ppb_kint_table}). 
The configurational binding volume $v_0$ is fixed to $0.80$~M$^{-1}$, the mean of the binding volumes obtained for G$_2$ and G$_4$-dPGS from our previous simulations.~\cite{xu:biomacro}
}
\end{figure*}

Corresponding to the fitting of binding coverages $\ti$ on G$_2$-dPGS and G$_4$-dPGS binding sites, as shown in Fig.~\ref{divalent}, the resulting $\dgb$ values are calculated as $-2.73\,\kt$ (G$_2$) and $-2.98\,\kt$ (G$_4$) for the MMvH model, whereas $-1.77\,\kt$ (G$_2$) and $-1.98\,\kt$ (G$_4$) for the PPB model. 
Table~\ref{mvh_ppb_kint_table} shows the values of $\dgb$ averaged over G$_2$-dPGS and G$_4$-dPGS cases.
It can be observed that both $\dgb$ and $\dga$ values from the MMvH model exceed (in magnitude) those from the PPB model across the whole $\ctmg \sim 0-25$\,mM range. This can again be attributed to higher approximations in the electrostatic partition coefficient designed in the MMvH model, based on the Debye--H\"{u}ckel charging free energy, as compared to that from the PPB model, incorporating non-linear effects in the electrostatic potential near the macromolecule vicinity.  
The standard intrinsic chemical potentials $\dgi^0$ after fitting the MMvH model $\ti$ with those from simulations are also given in Table~\ref{mvh_ppb_kint_table}.

Unlike the other models, we simultaneously fit both $\dga$ and $\dgb$ to perform numerical fitting of $\ta$ and $\tb$ obtained from the DM model with the simulation data. 
As shown in Table~\ref{mvh_ppb_kint_table}, the values of $\dga$ and $\dgb$ for the model turn out large and positive compared with those from other models, since the DM model tries to neutralize the entire dPGS charge via the electroneutrality condition in the dPGS phase.
The DM fits for $\ta$ differ to an extent with those from simulations, while those for $\tb$ are found to be reasonably good.   
The DM provides much better fits for $\ta$ in the case of G$_4$-dPGS as compared to G$_2$-dPGS. 
This is attributed to the bigger size of G$_4$-dPGS, which better satisfies the criterion $\kappa \rd \gg 1$, under which the DM electroneutrality condition holds comparatively well.

Having established the model frameworks by informing $\dgi$ by fitting the coverages $\ti$ to those from simulations and averaging the values of obtained $\dgi$ over generations (See Table~\ref{mvh_ppb_kint_table}), we finally utilize their predictive ability to explore the electrostatic characterization of dPGS for different generations and salt concentrations. 
As an example, Fig.~\ref{prediction}(d) shows the MMvH model predictions for the binding coverages $\ta$ and $\tb$ for the case of a competitive ion binding on G$_6$-dPGS, similar to Fig.~\ref{divalent} on G$_2$-dPGS and G$_4$-dPGS.
We also study the effective charge valency $\zeff$ of dPGS along with the composition of condensed ions on the molecule. 
Figs.~\ref{prediction}(a) and \ref{prediction}(b) show the variation of the effective charge valency $\zeff$ of G$_2$-dPGS and its normalized form $\zeff/\zd$, respectively, as a function of the DC concentration $\ctmg$, as predicted by the MMvH model. 
It can be clearly seen from Fig.~\ref{prediction}(a) that the introduction of DCs leads to a net charge renormalization of dPGS, which further decreases its $\zeff$. 
The inset shows that, with reference to the monovalent limit, the dPGS effective charge is $30-35\%$ further renormalized upon introducing DCs in the range of $1-4$\,mM, which is close to the physiological concentration range for calcium(II) ions. 
Fig.~\ref{prediction}(b) shows that the fraction of the bare dPGS charge that gets renormalized increases with the dPGS generation. The inset shows the variation for $\ctmg$ varying from $0$\,mM to $10$\,mM. 
The rate of dPGS charge renormalization with respect to $\ctmg$ is the highest at the low $\ctmg$ regime and subsides as $\ctmg$ increases, since the charge renormalized dPGS results in lower electrostatic binding chemical potential $\dgiel$. The reduced amount of renormalization is not attributed to the ion packing, which is evident from Fig.~\ref{prediction}(c) showing the total number of condensed ions (including both DCs and MCs) per dPGS sulfate group. 
As $\ctmg$ increases, the total number of condensed ions decreases, indicating that the ion packing effects diminish as $\ctmg$ increases. The decrease in the amount of renormalization thus predominantly has electrostatic origin.
Fig.~\ref{prediction}(a) shows that $80-90\%$ of the dPGS bare charge is renormalized as $\ctmg$ increases from $0-100$\,mM, however, the total number of condensed counterions effectively decreases, according to Fig.~\ref{prediction}(c). 
This in effect would significantly hamper the binding affinity of protein with dPGS. 
It has been well established through our previous works that the dPGS--protein complexation is dominantly influenced by the release of a few MCs that were highly confined due to strong charge renormalization.~\cite{xu:biomacro} 
The introduction of DCs, however, decreases the confinement of these condensed counterions, thus less counterions to be released during dPGS-protein binding. 
In addition, the strongly charge renormalized dPGS leads to lower electrostatic contribution to its overall binding affinity with the protein or any other multivalent ligand.

\section{Conclusion}
In this paper, we have addressed the biologically and industrially relevant problem of the competitive sorption  of mono- and divalent counterions into a highly charged globular polyelectrolyte, with direct comparison to CG simulations of the dendritic macromolecule dPGS. 
Beyond simple Donnan and ion-specific penetrable PB models, we introduced a  two-state discrete binding site model (MMvH) applicable for heterogeneous ligand systems (counterions with mixed valencies/stoichiometries).  
The broad classification of surrounding counterions as ``bound" and ``free" gives the MMvH model a computationally unique advantage over the PPB model, which involves the calculation of the distance-resolved counterion density profiles.  
The fitting results with simulations highlight the key differences in the MMvH and PPB models.  
Although being on a mean-field level, the PPB model incorporates non-linear electrostatic effects, which become more prominent near the surface of dPGS, delivering a relatively accurate picture of the dPGS--counterion electrostatic binding affinity, compared to the MMvH model, which approximates dPGS--counterion electrostatic interaction on a linearized PB (DH) level by absorbing these non-linear electrostatic effects into the effective charge valency $\zeff$ of dPGS. 
On the contrary, the MMvH model provides more accurate values of the extent of counterion adsorption $\Theta$ at high concentrations (\textit{i.e.}, in the binding site saturation regime) than the PPB model.
The reason is that the MMvH model assumes discrete binding sites, whereas the PPB model treats dPGS charge as continuum and allows an unlimited uptake of counterions, which is not realistic. 

Future extensions of the MMvH model could include an extra level of competition between adsorbed ions explicitly, namely through a non-linear term in Eq.~\eqref{fmx} (of the type used in the regular solution theory or the Flory--Huggins approximation in polymer theories) that describes the interaction between two adsorbed ions in proximal positions (sites). The effects of this generalization in a different context can be found in a study on ion induced lamellar-lamellar phase transition in charged surfactant systems.~\cite{Harries2006} In general, this type of competition results in non-continuous adsorption equilibria and could be interesting in the present context.

The simplest presented model, the Donnan model (DM) extended for ion-specific effects, is also useful for a quick, qualitative prediction of the adsorption ratio. 
Per construction it should become more accurate for large globules and/or large salt concentrations (for which the globule size becomes larger than the DH screening length), where the electroneutrality condition is better justified. 

The models presented in this work can be used to accurately extrapolate and predict the competitive ionic sorption in experiments for a wide range of salt concentrations and salt compositions. They can be also easily generalized to more ionic components and valencies.  
The electroneutrality radius required for the DM model and the intrinsic macromolecular charge distribution required for the PPB model as an input parameter (in the form of the bare radius $\rd$), are taken from simulations. 
However, they can also be derived by measuring the form factors from, \textit{e.g.}, neutron scattering.~\cite{Boris1996,Berndt2006}
The MMvH (ML) model requires the effective radius $\reff$ of the macromolecule as an input parameter, which besides simulations, can also be derived from independent experiments such as electrophoresis and fitting structure factors (of non-dilute colloidal suspensions) by DLVO interactions.~\cite{hunter,israelachvili2011intermolecular} 
As we showed, $\reff$ can also be obtained using PB models and related theories provided the intrinsic macromolecular charge distribution is available. 

\begin{acknowledgements}
The authors are indebted to Matthias Ballauff for insightful discussions. R.N. thanks Jacek Walkowiak for fruitful discussion. This project has received funding from the European Research Council (ERC) under the European Union's Horizon 2020 research and innovation programme (grant agreement No.\ 646659). X. X. acknowledges the National Science Foundation of China (21903045) and China Postdoctoral Science Foundation (2019M661842) for financial support. M.K. acknowledges the financial support from the Slovenian Research Agency (research core funding No.\ P1-0055).
\end{acknowledgements}

\section*{Data availability}
The data that support the findings of this study are available from the corresponding author upon reasonable request.

\bibliographystyle{aip}

\end{document}